\documentclass[10pt,twocolumn,superscriptaddress,aps,prx,balancelastpage,longbibliography,nofootinbib]{revtex4}
\usepackage[latin9]{inputenc}
\usepackage{color,multirow}
\usepackage{verbatim}
\usepackage{amsmath}
\usepackage{mathtools}
\usepackage{amssymb,ulem,manfnt}
\usepackage{graphicx,cellspace,booktabs}
\usepackage{esint}
\usepackage[unicode=true,pdfusetitle,
 bookmarks=true,bookmarksnumbered=false,bookmarksopen=false,
 breaklinks=false,pdfborder={0 0 1},backref=false,colorlinks=true]{hyperref}
\usepackage{breakurl}
\usepackage{epstopdf}
\usepackage{yfonts}
\usepackage[dvipsnames]{xcolor}

\makeatletter

\@ifundefined{textcolor}{}
{%
 \definecolor{BLACK}{gray}{0}
 \definecolor{WHITE}{gray}{1}
 \definecolor{RED}{rgb}{1,0,0}
 \definecolor{GREEN}{rgb}{0,1,0}
 \definecolor{BLUE}{rgb}{0,0,1}
 \definecolor{CYAN}{cmyk}{1,0,0,0}
 \definecolor{MAGENTA}{cmyk}{0,1,0,0}
 \definecolor{YELLOW}{cmyk}{0,0,1,0}
}

\usepackage[caption=false]{subfig}
\usepackage{bm}

\def\l@subsubsection#1#2{}
\makeatother
\begin{document}
\title{Extracting Wilson loop operators and fractional statistics from a \\ single bulk ground state}
\author{Ze-Pei Cian}
 \affiliation{Department  of  Physics,  University  of  Maryland,  College  Park,  Maryland  20742,  USA}
\affiliation{Joint Quantum Institute, College Park, 20742 MD, USA}
\author{Mohammad Hafezi}
\affiliation{Department  of  Physics,  University  of  Maryland,  College  Park,  Maryland  20742,  USA}
\affiliation{Joint Quantum Institute, College Park, 20742 MD, USA}
 \affiliation{Department of Electrical and Computer Engineering,  University  of  Maryland,  College  Park,  Maryland  20742,  USA}
\author{Maissam Barkeshli}
\affiliation{Department  of  Physics,  University  of  Maryland,  College  Park,  Maryland  20742,  USA}
\affiliation{Joint Quantum Institute, College Park, 20742 MD, USA}
\affiliation{Condensed Matter Theory Center, Department of Physics, University of Maryland, College Park, 20742 MD, USA}

\begin{abstract}
An essential aspect of topological phases of matter is the existence of Wilson loop operators that keep the ground state subspace invariant. Here we present and implement an \it unbiased \rm numerical optimization scheme to systematically find the Wilson loop operators given a single ground state wave function of a gapped Hamiltonian on a disk. We then show how these Wilson loop operators can be cut and glued through further optimization to give operators that can create, move, and annihilate anyon excitations. We subsequently use these operators to determine the braiding statistics and topological twists of the anyons, yielding a way to fully extract topological order from a single wave function. We apply our method to the ground state of the perturbed toric code and doubled semion models with a magnetic field that is up to a half of the critical value. From a contemporary perspective, this can be thought of as a machine learning approach to discover emergent 1-form symmetries of a ground state wave function. From an application perspective, our approach can be relevant to find Wilson loop operators in current quantum simulators.
\end{abstract}

\maketitle

\section{Introduction}

Topologically ordered phases are gapped quantum phases of matter that cannot be characterized by local order parameters, but rather by long-range entanglement and fractional statistics of quasiparticle excitations. For decades, a major question has been how to properly diagnose and characterize topological order in a quantum many-body system. While much progress has been made \cite{wen04,nayak2008,wang2008,senthil2015,zeng2019,barkeshli2019,kawagoe2020microscopic,bulmashSymmFrac,aasen2021characterization}, an outstanding question remains: Can we fully extract topological order from a single bulk ground state wave function, with no access to the Hamiltonian?

\begin{figure}[h]
\includegraphics[width=0.5\textwidth]{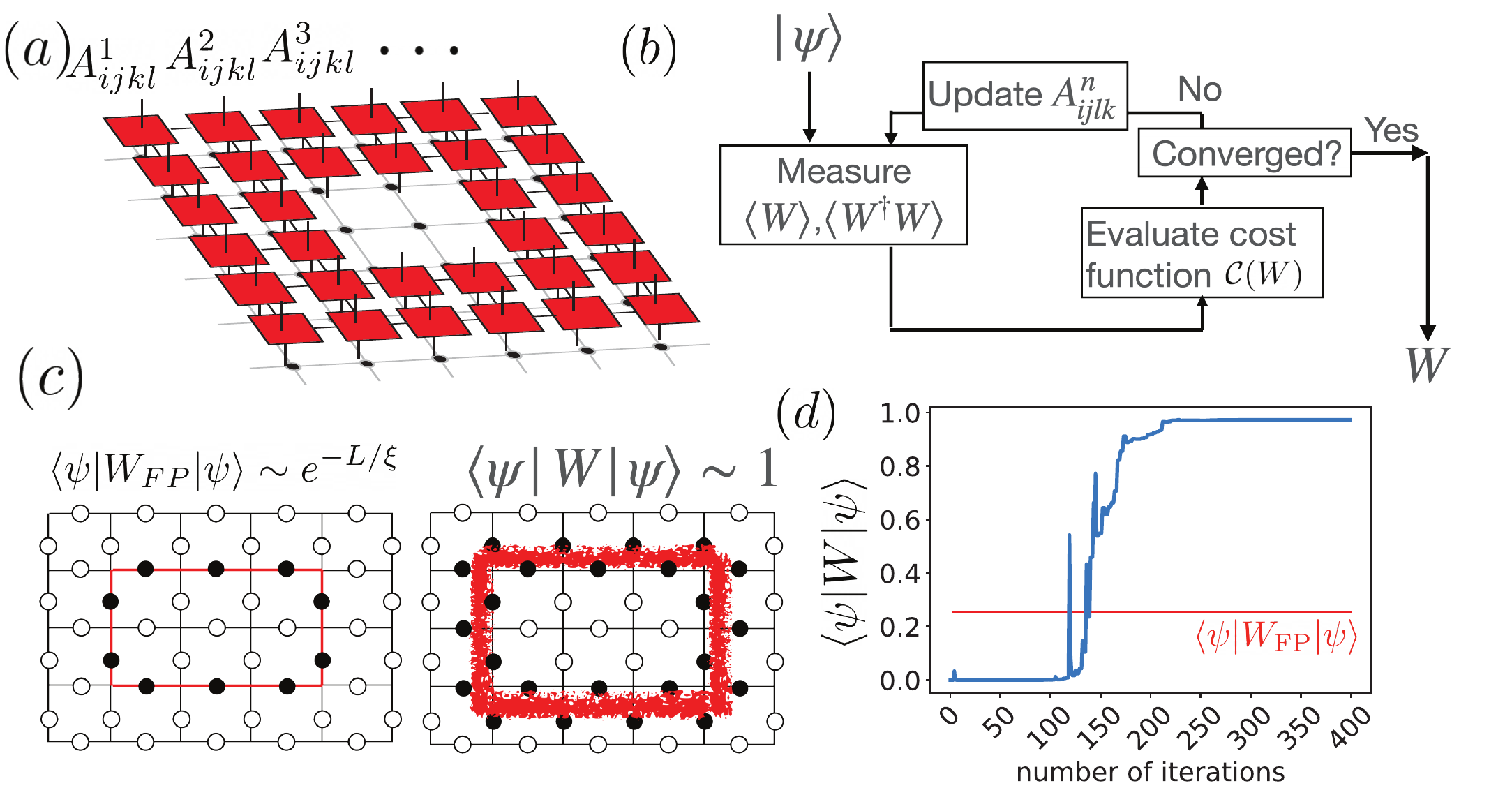}
    \caption{ (a)  WLO parameterized with a matrix product operator. (b) Numerical procedure to optimize a WLO.  (c) The expectation value of the WLO $W_{\rm FP}$ for the exactly solvable fixed point follows perimeter law when evaluated in the perturbed ground state: $\langle \psi | W_{\rm FP} | \psi \rangle \sim e^{- L/\xi}$, where $L$ and $\xi$ are the perimeter of $W_{\rm FP}$ and correlation length, respectively. However, the expectation value of the optimized Wilson loop operator does not decrease exponentially with the perimeter. (d) The expectation value of the Wilson loop operator $\langle \psi | W |\psi\rangle$ during the optimization iteration described in (b). The typical total number of iterations is around $400$. $|\psi\rangle$ is a ground state of the perturbed toric code model with $h_x = 0.15$, $h_z = 0.05$. The Wilson loop operator is a rectangle with side length $L_x =36$ and $L_y = 6$ and thickness 1.}
      \label{fig_1}
\end{figure}

Apart from the fundamental interest in the above question, there is a growing body of experimental effort in creating topologically ordered matter in quantum simulators. Recent examples include the implementation of toric code in superconducting qubit systems \cite{Googletoric2021}, and dimer models in Rydberg arrays \cite{HarvardQSL1,bluvstein2021quantum}. Since various kinds of perturbation are present in any experimental implementation, the precise Hamiltonian may not be known and may depart significantly from that of the pristine, idealized models. It is thus important to find a systematic approach to characterize topological order given a wave function, with minimal knowledge of the Hamiltonian.

%the realized Hamiltonian and associated WLOs can take different forms from that of the pristine, idealized models \cite{Hastings1D}, as schematically shown in \ref{fig_1}(a). For example, in recent Rydberg experiments the expectation value for the naive form of the Wilson loop operator (that of the fixed point) is far less than unity \cite{HarvardQSL1}. Therefore, it is important to find a systematic approach to extract WLOs given a wave function, with minimal knowledge of the Hamiltonian.

From a modern perspective, one key aspect of topological order is the existence of an emergent, higher symmetry. To each curve in space, there exists a set of Wilson line operators (WLOs), which correspond to adiabatically transporting topologically non-trivial quasiparticles along $\gamma$. If $\gamma$ is a contractible loop, the corresponding Wilson loop operators, or closed WLOs, keep a particular ground state invariant \cite{hastings2005quasiadiabatic}, while a WLO with open ends creates quasiparticle excitations near the two endpoints of $\gamma$. In this sense, the closed WLOs on contractible loops can be thought of as an \it emergent \rm symmetry of the ground state.\footnote{Closed WLOs on non-contractible loops can be thought of as a spontaneously broken emergent symmetry \cite{gaiotto2014}, because while they keep the ground state subspace invariant, implying an emergent symmetry, they act non-trivially on ground states, implying `spontaneous symmetry breaking.' } The symmetry is emergent in general because, aside from certain exactly solvable models \cite{kitaev2003,levin2005string}, the Hamiltonian need not commute with these WLOs. In contrast to ordinary symmetries, which are implemented by operators with support over the entire space, the closed WLOs have support only on loops; in the case where all topological quasiparticles are Abelian, the WLOs can be thought of, in modern terminology, as emergent 1-form symmetries of the system \cite{gaiotto2014}. These WLOs should in principle contain all of the data that characterizes the topological order, however it is not well-understood how to tease it out in practice. 
%Therefore, a fundamental question arises: is it possible to find these emergent symmetries, given a wavefunction with no access to the microscopic Hamiltonian?

In this work, we propose a numerical method to systematically search for a complete set of Wilson loop operators for the case of Abelian topological orders, using only the ground state of a gapped Hamiltonian defined on a disk-like region of space. %\commentmb{Again, do we need translationally invariant} \ZP{Not for WLOs and S matrix. But we need some symmetry for T matrix. I guess we can remove the translation symmetry condition here but introduce it in T matrix section.}. 
We do this by considering a variational ansatz for Wilson loop operators in terms of a matrix product operator with support on a ribbon along $\gamma$, as schematically shown in Fig. \ref{fig_1}(a). We then set up a cost function in terms of the variational parameters of the WLOs. The minima of the cost function, which we numerically optimize for, gives the WLOs as diagrammatically shown in Fig. \ref{fig_1}(c). The obtained Wilson loop operator expectation value can reach close to unity after a few hundred iterations (Fig. \ref{fig_1}(d)). We emphasize that our procedure is \it unbiased \rm and assumes no prior knowledge of the form of the WLOs. 

Once the WLOs are obtained, we show how one can perform further optimization-based schemes to find operators that can create, move, and annihilate the anyons. Finally, we show that these operators can be utilized to extract the modular $S$ and $T$ matrices of an Abelian topological order, which gives a complete characterization of the topological order. In particular the $S$ and $T$ matrices encode all of the information about the fractional statistics. %\MH{if we have a schematic figure, we can refer to it here.}

We successfully demonstrate our numerical protocol in models with non-zero correlation lengths. For example, we show how one can extract the modular $S$ and $T$ matrices from only the ground states of the perturbed toric code and doubled semion models, with a Zeeman field that is up to half of the critical value. 

To date, several invariants of two-dimensional topologically ordered states have been shown to be obtainable from the ground state wave function through a variety of methods. This includes the total quantum dimension measured through topological entanglement entropy  \cite{kitaev2006topological, levin2006detecting}, the many-body Chern number and Hall conductance \cite{dehghani2021extraction, cian2021many,fan2022extracting}, various invariants of symmetry-protected topological (SPT) states \cite{Shiozaki2017,Shiozaki2018many,elben2020many}, and the chiral central charge \cite{tu2013,zaletel2013,li2008,kim2021modular, kim2021chiral}. 
The modular $S$ and $T$ matrices, which encode details of the fractional statistics of the quasiparticles, can, under certain conditions, be extracted from the full set of ground states on a torus \cite{zhang2012quasiparticle,zhang2015,wen1990naberry} or in the presence of twist defects \cite{zhu2020}, but not to date from a single ground state on a disk. 

We note that Ref. \cite{bridgeman2016, wahl2020local} also proposed to find WLOs through an optimization approach, by searching for WLOs that commute with the Hamiltonian. However generic systems are not expected to have WLOs that commute with the Hamiltonian; instead as discussed above WLOs only appear as emergent symmetries that keep the ground state subspace invariant. Our work, in contrast to Ref.\cite{bridgeman2016, wahl2020local}, uses only the ground state without requiring knowledge of the Hamiltonian. 

The paper is organized as follows. In Sec. \ref{sec_WLO_and_anyon_data}, we review the basic properties of Wilson line operators and the algebraic theory of anyon. In Sec. \ref{sec_optimization_scheme}, we provide the optimization scheme for probing closed Wilson loop operators. In Sec. \ref{sec_manipulation}, we propose the scheme to create, move, annihilate anyons and measure the topological twist. We present numerical simulations for abelian topological order models in Sec. \ref{sec_numeric}. Finally, we provide an outlook for future works in Sec. \ref{sec_summary_and_outlook}.

\section{Wilson loop operators and anyon data}
\label{sec_WLO_and_anyon_data}
In this section, we briefly review the basic properties of WLOs and the algebraic theory of anyons. Since our goal is to extract topological invariants from the bulk of the wave function, in this section we consider a two dimensional system on an infinite plane. 

The anyon theory consists of a collection of algebraic data that characterize the universal topological properties, namely the fusion and braiding properties, of the anyonic excitations of a many-body system. The precise mathematical framework is that of a unitary modular tensor category (UMTC). For reviews of UMTCs in the context of topological phases of matter, see for example Ref. \cite{kitaev2006,Bonderson07b,wang2008,barkeshli2019,kawagoe2020microscopic}. For a detailed discussion of how to relate the algebraic data of the UMTC to the microscopic properties of a quantum many-body system, see Ref. \cite{kawagoe2020microscopic}. 

The description that we provide below can be made exact in the context of exactly solvable models, such as the toric code and its generalizations, the quantum double and Levin-Wen models \cite{kitaev2003,levin2005string}. The effect of perturbations to these exactly solvable models can also be studied systematically, using quasi-adiabatic continuation \cite{hastings2005quasiadiabatic}. For chiral topological phases, such as fractional quantum Hall states or fractional Chern insulators, which have no description in terms of an exactly solvable model, it is expected that the same discussion applies, although it has not been explicitly studied outside of the context of field theory. 

\subsection{Anyons, Wilson line and loop operators}

Since the system has a finite correlation length, we can define states with quasiparticle excitations that are localized on the scale of the correlation length. We can then group the quasiparticles into topological equivalence classes: two quasi-particle excitations are equivalent if and only if there is a local operator that can convert one into the other. The different equivalence classes define a finite set of distinct anyon types, sometimes also referred to as superselection sectors or topological charges, $\{I, a, b,c,... \}$. The set of anyons contains the identity sector $I$, which corresponds to excitations that can be created by local operators. 

Since the anyonic excitations can be localized to within a correlation length of a particular point in space, we can consider a state with anyon type $a$ at position $x$, and denote it as $|a_{x}\rangle$. In defining $|a_x\rangle$, we assume that far away from $x$, on the scale of the correlation length $\xi$, $|a_x\rangle$ locally looks like the ground state. Furthermore, we assume any other non-trivial topological charges are infinitely far away and do not include them in labeling the state $|a_x\rangle$. Note that since $a$ refers to an equivalence class of excitations, there are many states that can be labeled as $|a_x\rangle$, so our choice is not unique. 

By construction, the expectation value of any local observable $O(x')$ satisfies $\langle a_{x}|O(x')| a_x\rangle = \langle GS| O(x') | GS\rangle$, as long as $x$ and $x'$ are far away from each other, $|x-x'| \gg \xi$, where $|GS\rangle$ is the ground state of the system and $\xi$ is the correlation length. The above equality holds up to $O(e^{-|x - x'|/\xi})$ corrections. Physically, this corresponds to the fact that the state has short-range correlations, so that a disturbance in the vicinity of $x$ has exponentially decreasing effects in the ground state beyond a correlation length. 

\begin{figure}[ht]
    \includegraphics[width=0.5\textwidth]{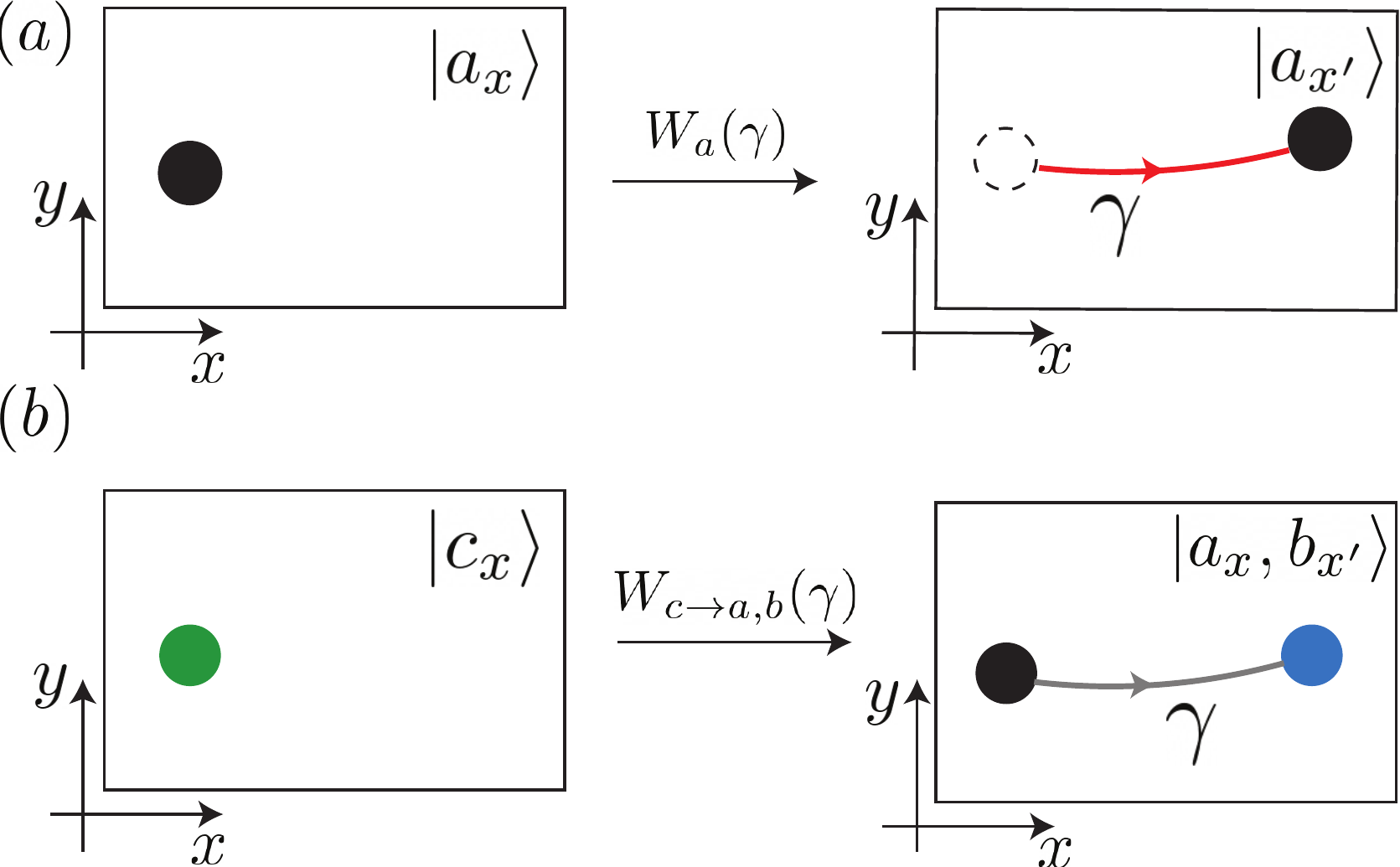}
      \caption{ (a) When the movement operator $W_a(\gamma)$ is applied to the state $|a_x\rangle$, the anyon excitation is moved to $x'$. (b) When the splitting operator $W_{c \rightarrow a,b}(\gamma)$ is applied to the state $|c_x\rangle$, the anyon splits into two anyons $a$ and $b$ at position $x$ and $x'$, respectively.  }
      \label{fig_move_split}
\end{figure}

The anyons can be moved from one place to another by applying an operator along an arbitrary path. In particular, we define a Wilson line operator $ W_a(\gamma)$ along a path $\gamma$ starting at $x_1$ and ending at $x_2$. $W_a(\gamma)$ moves the anyon excitation from position $x_1$ to $x_2$ along the path $\gamma$ as shown in Fig. \ref{fig_move_split}(a). Such operators are referred to as ``movement operators" in \cite{kawagoe2020microscopic}. Specifically,
\begin{eqnarray}
|a_{x_2}\rangle = W_a(\gamma) |a_{x_1}\rangle.
\end{eqnarray}
In general, $W_a(\gamma)$ has support, up to exponentially small corrections, on a ribbon of thickness on the scale of $\xi$, centered on $\gamma$. \footnote{Specifically,  $W_a(\gamma)$ can be approximated by a ribbon operator with finite thickness $t$ \cite{hastings2005quasiadiabatic}. The error of the approximate Wilson line operator $W$ is of order
$\epsilon = |\langle \psi | W - W_{\rm exact} |\psi \rangle| \sim O(N_s e^{-t/\xi})$,
where $N_s$ is the number of sites in the support of $W$, and $W_{\rm exact}$ is the exact, and presumably non-local WLO, for the ground state wave function $|\psi\rangle$.}
The precise choice of the operator $W_a(\gamma)$ is not unique in general,
and a precise definition is also non-universal and depends on the microscopic details of the system. Nevertheless, $W_a(\gamma)$ encodes certain universal topological data that we wish to extract. 

Note that $W_a(\gamma)$ in general need not be a unitary or even invertible operator, although in many simple examples, particularly when $a$ is an Abelian anyon, $W_a(\gamma)$ can be chosen to be unitary. Moreover, when $a$ is Abelian, we can take
\begin{align}
    W_a^\dagger(\gamma) = W_{\bar{a}}(\gamma) = W_a(-\gamma) ,
\end{align}
where $-\gamma$ refers to the path $\gamma$ traversed in the opposite direction and $\bar{a}$ is the anti-particle of $a$. That is, $W_a^\dagger(\gamma)$  effectively takes $\bar{a}$ along $\gamma$ or, equivalently, takes $a$ along the path $-\gamma$. 
%We can also consider the case where $\gamma$ is a closed loop. In this case, $W_a(\gamma)$ is referred to as a loop operator. In the limit 
%\begin{align}
%    W_a(\gamma) |GS \rangle = d_a |GS \rangle. 
%\end{align}
%Here $d_a \geq 1$ is the quantum dimension of the anyon $a$. Note that in principle one could absorb $d_a$ into the definition of 

We can also define loop operators by picking $\gamma$ in $W_a(\gamma)$ to be a closed loop. Physically this can be understood as creating $a$ and its dual $\bar{a}$ out of the ground state, moving one around the loop $\gamma$, and reannihilating. If $\gamma$ is a contractible loop in the space, such operators should keep the ground state invariant. Therefore, for each loop $\gamma$ we have a loop operator $W_a(\gamma)$, which keeps the ground state invariant:
\begin{align}
\label{loopEqn}
W_a(\gamma) |GS \rangle = d_a |GS \rangle . \end{align}
Here $d_a \geq 1$ is referred to as the quantum dimension of the anyon $a$, and is part of the universal data of the UMTC. We
have chosen a convention where $d_a \geq 1$ appears on the RHS; we could in principle absorb $d_a$ into the definition of $W_a(\gamma)$. The choice above allows us to make contact with the fusion algebra of the UMTC description. 

As with the line operators, the Wilson loop operators $W_a(\gamma)$ are in general not unitary operators, unless $a$ is an Abelian anyon, in which case we also have $d_a = 1$.

Eq. \ref{loopEqn} makes explicit that the ground state of a topologically ordered state has emergent symmetries, as there are loop operators that keep the ground state invariant. Importantly, the operators are supported, up to exponentially small corrections, on a codimension-1 region, and therefore do not correspond to ordinary global symmetries, which have support over the entire space. When the anyons are Abelian, the Wilson loop operators form a group structure and are referred to as 1-form symmetries \cite{gaiotto2014}; more generally they are referred to as categorical or non-invertible symmetries. 

\subsection{Fusion rules and splitting operators}

%In this section, we describe the properties of fusion rule.
The anyons define a fusion algebra
\begin{eqnarray}
a \times b = \sum_{c} N_{ab}^c c, 
\end{eqnarray}
where the fusion multiplicities $N_{ab}^c$ are non-negative integers, which indicate the number of different ways the anyons $a$ and $b$ can be fused to produce the anyon type $c$. Each anyon type $a$ has a unique anti-particle $\bar{a}$, where $\bar{a} \in \mathcal{C}$ is such that  $N_{a{\bar {a}}}^{I}\neq 0$. Note that we can define a fusion matrix $N_a$, with entries $(N_a)_{bc} = N_{ab}^c$; the quantum dimension $d_a$ is then the largest eigenvalue of $N_a$.

An anyon $a$ is Abelian if and only if it gives a unique fusion outcome upon fusing with another anyon $b$. That is, given $b$, $N_{ab}^c = 1$ for a unique $c$ and $N_{ab}^c = 0$ otherwise. 

The fusion rule leads to the following relation for the Wilson loop operator:
\begin{equation}
    W_a(\gamma) W_b(\gamma) |GS \rangle = \sum_c N_{ab}^c  W_c(\gamma) |GS \rangle,
\end{equation}
whenever $\gamma$ is a loop. Note that with the conventions chosen above, this implies $d_a d_b = \sum_c N_{ab}^c d_c$. 

In addition to movement operators and loop operators, we can define splitting operators. For simplicity, here we only introduce the spliting operator in the case $N_{ab}^c \leq 1$; the generalization can be found in Ref. \cite{kawagoe2020microscopic}. Suppose that $c$ is contained in the fusion outcome of $a$ and $b$, that is, $N_{ab}^c = 1$. 
We can define a splitting operator, 
\begin{eqnarray}
W_{c \rightarrow a,b}(\gamma)|c_{x_1}\rangle = |a_{x1}, b_{x_2}\rangle,
\end{eqnarray}
where $|a_{x1}, b_{x_2}\rangle$ denotes a state with two excitations: anyon $a$ at position $x_2$ and anyon $b$ at position $x_1$ and $|x_2 - x_1| \gg \xi$ as shown in Fig. \ref{fig_move_split}(b). 

One can also create the anyon $a$ and its antiparticle $\bar{a}$ by applying 
\begin{equation}
    W_{I \rightarrow \bar{a}, a}(\gamma)|I_{x_1}\rangle = |\bar{a}_{x_1}, a_{x_2}\rangle,
\end{equation}
where $|I\rangle$ is a state in the identity superselection sector.  

Observe that if we start with a loop operator $W_a(\gamma)$, and we project part of the operator along some segment of $\gamma$ to the identity, then we obtain a cut operator that effective creates an anyon and its anti-particle out of the vacuum. Therefore we can obtain a choice of $W_{I \rightarrow a, \bar{a}}(\gamma_{\text{cut}})$ by starting with $W_a(\gamma)$ for a loop $\gamma$ and implementing the above cutting procedure.

\subsection{Modular $S$ matrix and twist product}
\label{sec_Braiding_statistic}

A large portion of the universal data of a topological phase of matter is encoded in the modular $S$ and $T$ matrices. In fact for almost all topological phases of interest in physics, the $S$ and $T$ matrices provide a complete set of invariants. 

The modular $S$ matrix contains information about the mutual braiding statistics between far separated anyon excitations, and also completely defines the fusion coefficients $N_{ab}^c$. 

In particular, $S_{ab}$ is the quantum mechanical amplitude of the process where a particle of type $a$ and another particle of type $b$ are created and separated, the particle $a$ is moved around the particle $b$, and then the particle-anti-particle pairs are annihilated.

%The modular $S$ matrix carries the dynamical information from braiding anyons in $2+1$ dimensions. An equivalent braiding matrix $\tilde{S}$ can also be measured through the ground state wave function in the $2+0$ dimension space. 

\begin{figure}[h]
    \includegraphics[width=.48\textwidth]{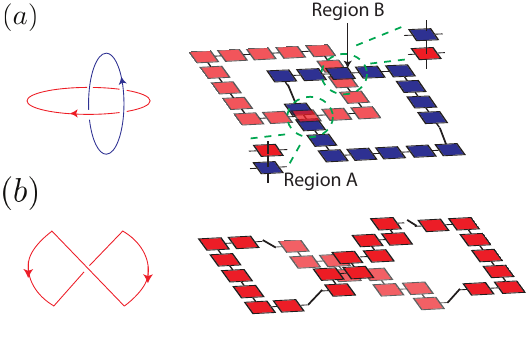}
      \caption{(a) {\bf Mutual-braiding statistics.} To measure the mutual braiding statistics, we calculate twist product for two Wilson loop operators $W_a$ and $W_b$. The supports of the two Wilson loop operators are divided into two regions. In region A, the order of the product is $W_a W_b$ and in the region B, the order of the product is reversed. (b) The matrix product operator that measures the self-braiding statistic $T_a$. %\commentmb{why do the arrows look like this?}
      }
      \label{fig_S}
\end{figure}

Given a set of closed Wilson loop operators $W_a$, where $a \in \mathcal{C}$, we can define a matrix $\tilde{S}_{ab}$, as
\begin{eqnarray}
\tilde{S}_{a,b} = \frac{\langle GS| W_a \infty W_b |GS \rangle}{\langle GS | W_a W_b |GS \rangle},
\end{eqnarray}
where $\infty$ is the twist product (see e.g. Ref. \cite{haah2016invariant}) and shown in Fig. \ref{fig_S}. For arbitrary WLOs $P$ and $Q$ defined in regions $A$ and $B$ as shown in Fig. \ref{fig_S}(a), $P = \sum_{k} P^A_k \otimes P^B_k$, $Q = \sum_{l} Q^A_l \otimes Q^B_l$, the twist product $P\infty Q$ is defined as
\begin{eqnarray}
P \infty Q = \sum_{kl} P_k^A Q_l^A \otimes Q_l^B P_k^B,
\label{Twist product}
\end{eqnarray}
where the product order is reversed in the region $B$.

Note that to define the above twist product, we need each operator $W_a$ to have support strictly on a ribbon of finite thickness.

We expect that $\tilde{S}_{ab}$ is related to $S_{ab}$ as 
\begin{align}
    \tilde{S}_{ab} = \frac{d_a d_b}{\mathcal{D}} S_{ab},
\end{align}
where $\mathcal{D} = \sqrt{\sum_a d_a^2}$ is the total quantum dimension. In this paper we only work with Abelian anyons, in which case $d_a = 1$ and $\mathcal{D}^2$ is the total number of distinct anyon types. 

For Abelian anyons, the braiding phase between anyon $i$ and $j$ can be measured from the phase of the twist product 
\begin{eqnarray}
\phi_{i,j} = {\rm arg}[\tilde{S}_{i,j}].
\label{braiding_phase}
\end{eqnarray}

\subsection{Modular $T$ matrix}

The modular $T$-matrix is a diagonal matrix, 

\begin{eqnarray}
    T_{ab} = \theta_a \delta_{ab}
    \label{eq_T_mat}
\end{eqnarray}
where $\theta_a$ is the topological twist of the anyon $a$. Due to the spin-statistics theorem, $\theta_a$ also corresponds to the exchange statistics of $a$. In order to exchange a pair of identical anyons, we first create two anyon and anti-anyon pairs from the ground state, and we then move the two identical anyons and exchange them. Finally we fuse the anyon and anti-anyon and return the the ground state  as shown in Fig. \ref{fig_S}(b). If we normalize the process properly, the net effect of this procedure gives the exchange statistics of the anyons. 

One can create and exchange anyons and measure the exchange statistics using the Wilson loop operators. The detailed implementation of the extraction of exchange statistics using the Wilson loop operators is given in Sec. \ref{sec_twist}.

%\subsection{Local invisibility of the Wilson loop operators}

%If the path $\gamma$ forms a closed loop, one can treat the closed Wilson loop operators as a creation of anyon and anti-anyon pairs and the anyon travels around the path $\gamma$ and subsequently annihilate with the anti-anyon. This procedure does not create any extra excitation and therefore it keeps the ground state subspace invariant. Therefore, for an arbritrary closed Wilson loop operators, we have
%\begin{eqnarray}
%W_a(\gamma)|GS\rangle = d|GS\rangle,
%\end{eqnarray}
%where $d$ is a complex number.

%The local invisibility indicates that the ground state of the system has emergent symmetries \cite{gaiotto2014}. 
%Moreover, if the ground state wave function has multiple locally invisible operators and they do not commute with each other,it can be shown that the wave function can not be prepared through constant depth circuit from a trivial product state. Therefore, the local invisibility is also an multipartite entanglement witness \cite{haah2016invariant}.

\section{Optimization Scheme}

\label{sec_optimization_scheme}
\begin{figure}[t]
    \includegraphics[width=0.5\textwidth]{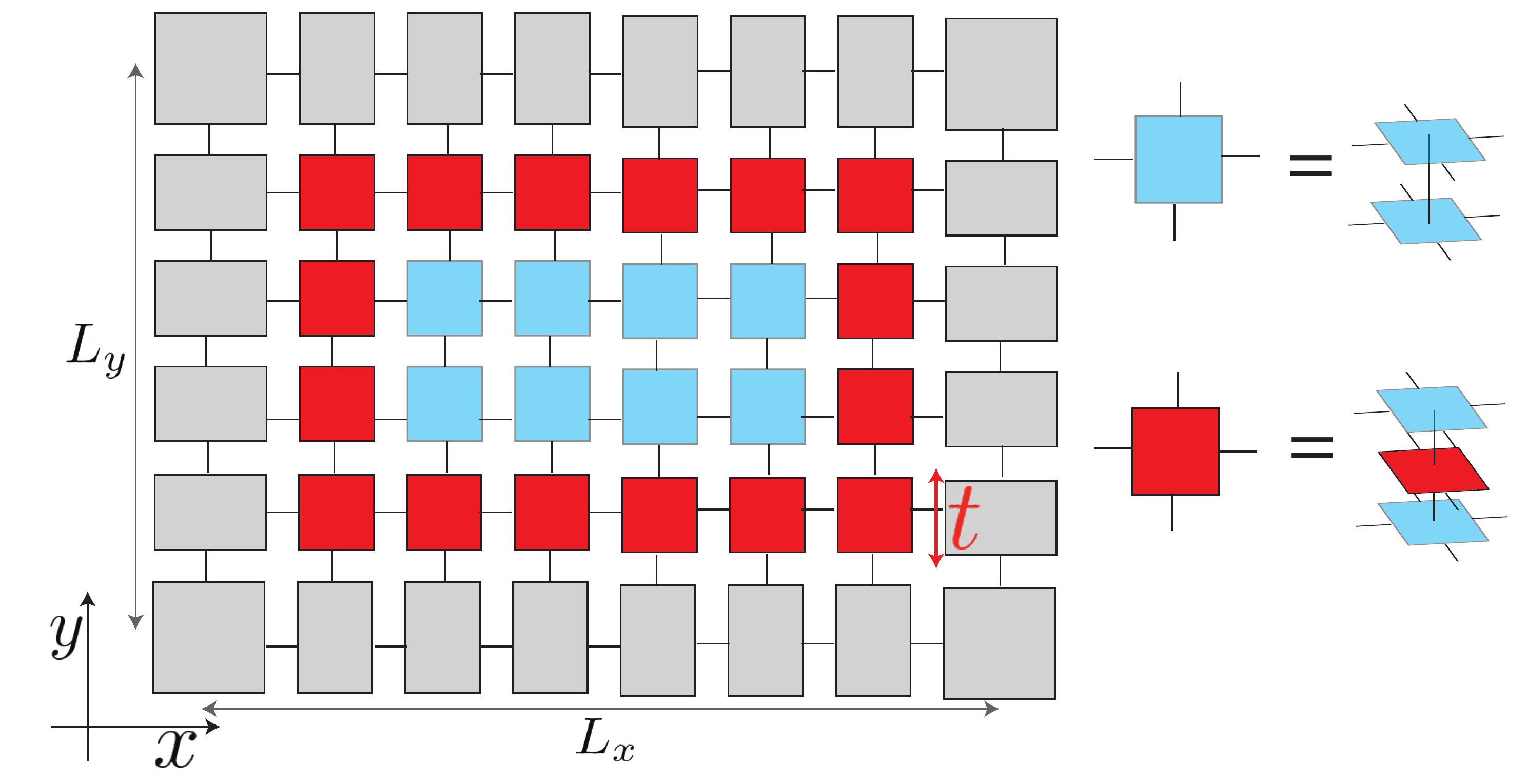}
      \caption{Illustration of the tensor contraction for calculating $\langle \psi | W | \psi \rangle$. The gray tensors represents corner transfer matrix (CTM). The blue tensors are the contraction of the bra and ket of the PEPS wave function.  The two are connected by contraction of the physical bond. The red tensors are the contraction of the PEPS wave function and the WLOs. }
      \label{fig_peps}
\end{figure}

%WLOs are one-dimensional like operators that can create and move anyons in the two-dimensional topological ordered system. A WLO with open ends creates an anyon and anti-anyon pair exciations localized at ends of the WLO. They can be annihilated with each other and form a closed loop as the anyon and anti-anyon meet. If the closed loop is a non-contractible loop on a torus, it can map the ground states of the topological ordered phase from one to another, and therefore it can serve as a logical operation for topological quantum computation \cite{nayak2008non}. However, if it is contractible, it leaves the ground state wave function invariant.

In order to study the WLOs in the bulk of a ground state wave function, we propose a numerical scheme to extract contractible closed WLOs  $W_a (\gamma)$ as defined in Eq. \eqref{loopEqn}. We parametrize the WLOs by an ansatz based on matrix product operators (MPOs) \cite{verstraete2008matrix, bridgeman2016,bultinck2017anyons}. 
The ansatz is defined by two parameters: $(R, \chi)$, where $R$ is a set of sites corresponding to the support of the WLO and $\chi$ is the bond dimension, as shown in Fig. \ref{fig_1}. For a certain class of analytically solvable topologically ordered states, e.g. the Levin-Wen model and the Kitaev quantum double model, it is known that the WLOs can be efficiently expressed by MPOs with a support region $R$ with small thickness $t$ and bond dimension $\chi$ \cite{kitaev2003,levin2005string, bultinck2017anyons}. 

%We note that the computational complexity grows up exponentially with the thickness $t$ to contract tensors and compute the expectation value $\langle \psi | W |\psi\rangle$. Therefore, it can be potentially  

%\commentmb{Make subsequent discussion consistent with the existence of Sec. II}
To extract the closed WLOs using the MPO ansatz, we numerically optimize the MPO ansatz in order to obtain a $W_a(\gamma)$ which approximately satisfies Eq. \eqref{loopEqn}.  In order to efficiently optimize a Wilson loop, we note that for Abelian topological orders, {\it an operator $W$ and a wave function $|\psi\rangle$ satisfies Eq. \eqref{loopEqn} if and only if 
\begin{eqnarray}
\langle \psi | W |\psi \rangle = 1, 
\label{wilson loop condition expectation value 1}\\
\langle \psi | W^\dagger W |\psi \rangle = 1.
\label{wilson loop condition expectation value 2}
\end{eqnarray}
}
The forward proof is trivial. For the backward proof, we assume that $W$ and $|\psi\rangle$ satisfy Eqs. \eqref{wilson loop condition expectation value 1} and \eqref{wilson loop condition expectation value 2} and without loss of generality, $W|\psi\rangle = a|\psi\rangle + b|\psi_{\perp}\rangle$, where $a$ and $b$ are complex numbers and $\langle \psi | \psi_{\perp}\rangle = 0$. Solving Eqs. \eqref{wilson loop condition expectation value 1} and \eqref{wilson loop condition expectation value 2} we can obtain $a = 1$ and $b = 0$.
Note that Eq. (11) implies that $a=1$ and Eq. (12) ensures that $W |\psi\rangle$ is normalized to $1$, which then requires $b = 0$. 

Therefore, we can define the cost function for a wave function $|\psi\rangle$ as 
\begin{eqnarray}
\mathcal{C}(W) = [\langle \psi | W |\psi \rangle -1]^2 + [\langle \psi | W^\dagger W |\psi \rangle -1]^2.
\label{cost}
\end{eqnarray}
It reaches a global minimum $\mathcal{C}(W) = 0$ only when $W$ is an exact Wilson loop operator. 

To variationally optimize the WLO, we start by initializing a random MPO with fixed $(R, \chi)$. Each tensor in the MPO is initialized randomly and independently from each other. For a translationally invariant system, one may naively expect that a translation symmetric closed WLO is a better ansatz. However, we found that the translation symmetric ansatz tends to be unstable numerically, leading to diverging or vanishing gradients in the optimization procedure. 

After the initialization, we minimize the cost function defined in Eq. \eqref{cost} through gradient based optimization. In this work, we apply the Adam algorithm \cite{kingma2014adam} to minimize the cost function. In this work, we fix the hypermeters of the Adam algorithm as $\beta_1 = 0.9, \beta_2 = 0.999$ and learning rate $10^{-3}$.
We iterate the optimization procedure until the cost function converges, which typically takes a few hundred to a few thousand iterations. 

We repeat the initialization and minimization $N_{\rm sample}$ times to obtain $N_{\rm sample}$ optimized WLOs $W_k$, where $1 \leq k \leq N_{\rm sample}$. Throughout the manuscript, we fix $N_{\rm sample} = 20$. We then measure the braiding phases and topological twists to classify the WLOs through the equivalence relation described as follows : we compute the mutual-braiding phases between $W_i$ and $W_j$, $\phi_{i,j}$, where $1 \leq i,j \leq N_{\rm sample}$ and topological twist for $W_i$ $\theta_i$, where $1 \leq i \leq N_{\rm sample}$. We say the two WLOs $W_i$ and $W_j$ are equivalent when $\phi_{i,k} = \phi_{j,k}$ for $1\leq k \leq N_{\rm sample}$, and $\theta_{i}=\theta_j$.  After grouping the WLOs into equivalence classes, we randomly pick one representative WLO from each equivalent class.

This equivalence relation assumes that if two WLOs have the same braiding phase with the rest of the WLOs and identical topological twists, the two WLOs are equivalent. We note that this condition only holds when we obtain a complete set of WLOs in our optimization procedure. Missing one could result in a false classification. However, one can verify whether a complete set of WLOs is found by checking if the resulting $S$ matrix is a unitary matrix. If one or more WLOs are missing, we can vary the hyper-parameters such as increasing the thickness of the WLOs or the bond dimension $\chi$.

The bottleneck of the numerical optimization is in the tensor contraction when calculating the expectation values $\langle \psi| W | \psi\rangle$ and $\langle \psi | W^\dagger W | \psi\rangle$. Here, we briefly describe the tensor contraction scheme and its computational time complexity. Given a ground state wave function $|\psi\rangle$, represented by an infinite projective entangled pair state (iPEPS) \cite{orus2009simulation, liao2019differentiable, crone2020detecting} as shown in Fig.~\ref{fig_peps} , we use the following procedure to evaluate the expectation value $\langle \psi| W |\psi\rangle$. 
Consider a closed WLO that takes the form of a rectangular loop with side lengths $L_x$ and $L_y$ as shown in Fig. \ref{fig_peps}, for a system on a square lattice. We first contract all tensors at $x=0$ to form a tensor $M$ with $L_y$ bonds as shown in Fig.~\ref{fig_peps}. We then contract the tensors at $x=1$ with $M$ one by one from $y = 0$ to $y = L_y+1$. We then repeat this procedure for all $x \leq L_x+1$. In this contraction procedure, the computational cost for contracting the tensors scales linearly with the number of sites along the $x$ direction ($L_x$) and exponentially with number of sites along the $y$ direction ($L_y$). Therefore, the total computational cost is bounded by $O(L_x L_y \chi ^{L_y+5})$. Moreover, since the thickness $t$ and the size of the hole need to be much larger than the correlation length, we have that $L_x, L_y > 3 \xi$. Thus the total computational cost for calculating the expectation value scales up as $O(\xi^2 \chi^{\alpha \xi})$ for some constant $\alpha \gg 1$. Therefore, this method is particularly suitable for models with small correlation length. Importantly, the complexity of the computation scales with the correlation length, not the total system size. 

\section{Manipulation of anyons}
\label{sec_manipulation}

Once we have obtained the closed WLOs, we can extract many non-trivial properties of the anyons. In particular, we can further obtain operators that create, move, and annihilate anyons. 

In the following, we discuss how to manipulate anyons by starting from the closed WLOs. For simplicity in this section,  we assume that the thickness of WLOs is $t = 1$ . The idea can be easily generalized to $t > 1$ as is the case for our numerical results in Section \ref{sec_numeric}.

\subsection{Removing and adding a virtual bond}

%Removing a virtual bond on a rank-$n$ tensor allows creating boundaries from in a closed WLO 
Before we proceed to the manipulation of anyons, we first define two basic operations of a tensor $A$, which is that of removing and adding a virtual bond. These two operations will be extensively used throughout this section. 

Removing a virtual bond is useful for cutting open closed WLOs. To remove a virtual bond in a tensor $A_{\alpha, \beta, \dots, \gamma}$, we define an edge tensor $E_\alpha$ that describes the boundary condition and contract the edge tensor $E_\alpha$ with $A_{\alpha, \beta, \dots, \gamma}$; the resulting tensor is 
\begin{eqnarray}
A'_{\beta, \dots, \gamma} = \sum_{\alpha} E_\alpha A_{\alpha, \beta, \dots, \gamma},
\end{eqnarray} 
and it has rank $n-1$.

Adding a virtual bond is useful when extending an open WLO or connecting two open WLOs. To add a trivial virtual bond to an arbitrary tensor $A_{\alpha, \beta, \dots, \gamma}$ with rank $n$, we define a new tensor $\tilde{A}$ with rank $n+1$ as
\begin{eqnarray}
\tilde{A}_{\alpha', \alpha, \beta, \dots, \gamma} = A_{\alpha, \beta, \dots, \gamma},
\end{eqnarray}
for all $1 \leq \alpha' \leq \chi$, where $\chi$ is the bond dimension of the new virtual bond.

Using the above two basic operations on tensors, we can create, move and annihilate anyons, as described in the following sections. 

\subsection{Creation of anyon and anti-anyon pairs}
\label{sec_Creation}

\begin{figure}[t]
    \includegraphics[width=.48\textwidth]{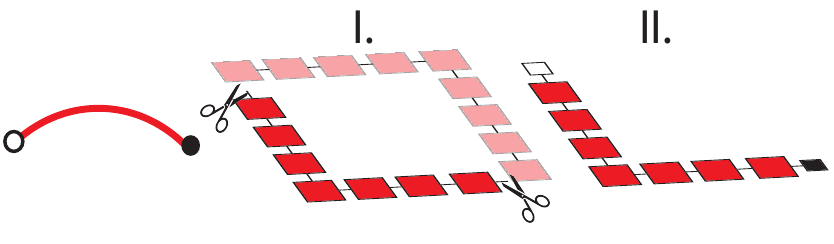}
      \caption{{\bf Creation of an anyon and anti-anyon pair.} {\bf Step I} : We truncate a closed WLO by discarding a segment of it. {\bf Step II} : We contract two two boundary tensor with edge tensor to terminate the open WLO.}
      \label{fig_open}
\end{figure}

To create an anyon and anti-anyon pair, we simply discard a segment of the closed WLO and apply an arbitrary edge tensor to terminate its boundary as shown in Fig. \ref{fig_open}. Specifically, we start with a closed WLO obtained using the procedure described in Sec. \ref{sec_optimization_scheme} for a give ground state wave function $|\psi\rangle$, which is  of the form
\begin{eqnarray}
W = \sum_{\{\alpha \}} A^1_{\alpha_0, \alpha_1} A^2_{\alpha_1, \alpha_2} \cdots A^{L-1}_{\alpha_{L-2}, \alpha_{L-1}} A^{L}_{\alpha_{L-1}, \alpha_0},
\end{eqnarray}
where $L$ is the length of the closed WLO.

We then keep the tensors from sites $1,\cdots, l$, as shown in step I of Fig. \ref{fig_open}, so that the new MPO is of the form
\begin{eqnarray}
\tilde{W}_{\beta, \beta'} = \sum_{\{ \alpha \}} A^1_{\beta, \alpha_1} A^2_{\alpha_1, \alpha_2} \cdots A^{l-1}_{\alpha_{l-2}, \alpha_{l-1}} A^{l}_{\alpha_{l-1}, \beta'}.
\end{eqnarray}

We then remove the virtual indices $\beta$ and $\beta'$ $\tilde{W}_{\beta,\beta'}$ by  applying the edge tensor $E$. The edge tensors can be chosen arbitrarily, since different choices are related to each other by a local operator. Throughout out this article, we use
\begin{eqnarray}
E_{\gamma} =
\begin{cases}
1, & \text{if}~\gamma = 1 \\
0, & \text{otherwise}
\end{cases}.
\end{eqnarray}

After removing the edge virtual bond shown in step II in Fig. \ref{fig_open}, the open WLO is of the form
\begin{eqnarray}
W^{\rm open} = \sum_{\beta, \beta'} \tilde{W}_{\beta, \beta'} E_\beta E_{\beta'}.
\end{eqnarray}

When another WLO passes through the open WLO, the system acquires an anyonic braiding phase as long as the crossing point is away from the boundary of the open WLO by an $O(\xi)$. 

Finally, in order to preserve the norm of the wave function, we normalize the open WLO defined above by a factor $\sqrt{\langle \psi | W^{\rm open \dagger}W^{\rm open}|\psi\rangle}$.

\subsection{Moving anyons}
\label{sec_Moving}

Let us imagine we have a WLO $W^{\text{open}}_{x_1, x_2}$ that creates an anyon $a$ at one endpoint $x_1$ and its conjugate $\overline{a}$ at the other endpoint $x_2$. We can use this to construct a different operator $W^{\text{open}}_{x_1', x_2}$, effectively moving $a$ from $x_1$ to $x_1'$. We can do this as follows. 

\begin{figure}[t]
    \includegraphics[width=.48\textwidth]{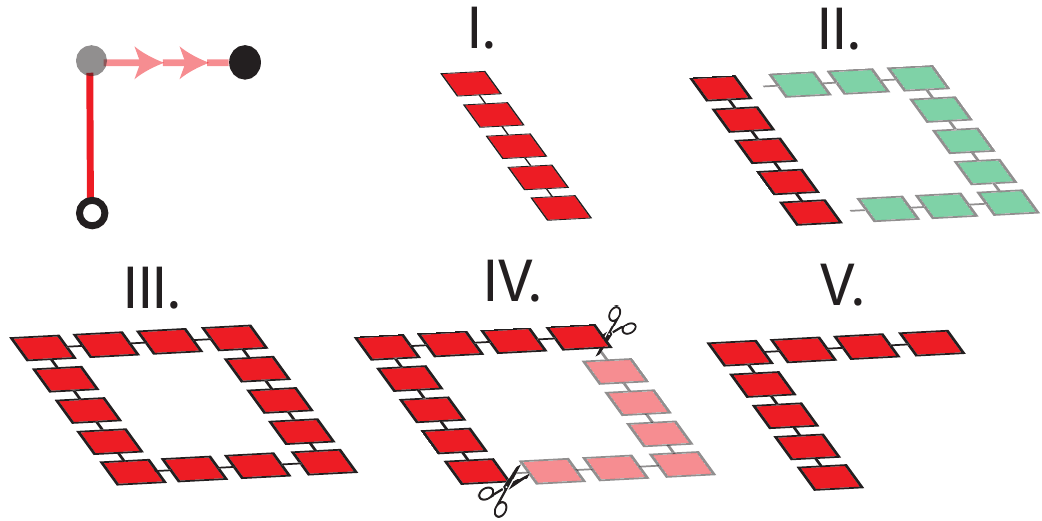}
      \caption{{\bf Moving anyons.} {\bf Step I} : We start with an open WLO that can be generated using the method described in Sec. \ref{sec_Creation}. {\bf Step II} : We randomly initialize an open MPO with virtual bond at the boundary tensors. The initialized open MPO and the open WLO form a closed MPO.  {\bf Step III}: We vary the closed MPO to minimize the cost function defined in Eq. \eqref{cost}. {\bf Step IV and V} Finally, we cut the closed WLO to create two excitations that can be located at different sites from the anyons in Step I. }
      \label{fig_move}
\end{figure}

We start with an open WLO as shown in step I of Fig. \ref{fig_move}. The open WLO with length $L$ is of the form
\begin{eqnarray}
W^{\rm open} = \sum_{\{\alpha\}}A^1_{\alpha_1} A^2_{\alpha_1,\alpha_2} \cdots A^{L-1}_{\alpha_{L-2},\alpha_{L-1}} A^{L}_{\alpha_{L-1}}.
\end{eqnarray}
The open WLO can be obtained using the procedure described in Sec. \ref{sec_Creation}.

We initialize another random MPO with length $L'$ that is a loop complement of the WLO $W^{\rm open}$ as shown in step II of Fig. \ref{fig_move} and is of the form
\begin{eqnarray}
W^{\rm open}_{C} = \sum_{\{\beta\}}C^1_{\beta_1} C^2_{\beta_1,\beta_2} \cdots C^{L'-1}_{\beta_{L'-2},\beta_{L'-1}} C^{L'}_{\beta_{L'-1}},
\end{eqnarray}

To connect $W^{\rm open}$ and $W^{\rm open}_P$ at the boundary, we add a trivial virtual bond on each boundary tensor. By adding trivial virtual bond on all the boundary tensors $A^1, A^{L}, C^1, C^{L'}$, we have
\begin{eqnarray}
\tilde{W}^{ \rm open }_{\gamma_0, \gamma_{L}} &=& \sum_{\{\alpha\}}\tilde{A}^1_{\gamma_0, \alpha_1} A^2_{\alpha_1,\alpha_2} \cdots A^{L-1}_{\alpha_{L-2},\alpha_{L-1}} \tilde{A}^{L}_{\alpha_{L-1}, \gamma_L}, \notag \\
\tilde{W}^{ \rm open }_{C, \mu_0, \mu_{L'}} &=& \sum_{\{\beta\}}\tilde{C}^1_{\mu_0, \beta_1} C^2_{\beta_1,\beta_2} \cdots C^{L'-1}_{\beta_{L'-2},\beta_{L'-1}} \tilde{C}^{L'}_{\beta_{L'-1}, \mu_{L'}}. \notag \\
\end{eqnarray}

We can therefore have a closed WLO of the form
\begin{eqnarray}
W = \sum_{\gamma_0, \gamma_L, \mu_0, \mu_{L'}}\tilde{W}^{ \rm open }_{\gamma_0, \gamma_{L}} \tilde{W}^{ \rm open }_{C, \mu_0, \mu_{L'}} \delta_{\gamma_0, \mu_{L'}} \delta_{\gamma_L, \mu_0},
\end{eqnarray}
where $\delta_{i,j}$ is a Kronecker delta function.

In step III of Fig. \ref{fig_move}, we minimize the cost function of Eq. \eqref{cost} for the closed WLO defined above. We note that in addition to vary the tensors in the $\tilde{W}^{\rm open}_{C, \mu_0, \mu_{L'}}$, we also have to vary tensors at the boundary of $\tilde{W}^{\rm open}_{\gamma_0, \gamma_L}$ i.e. $\tilde{A}^1$, $\tilde{A}^L$,...etc. in order to eliminate the anyon excitation at the boundary. 

We define a length parameter $b$, which is of order $O(\xi)$. In the optimization procedure, we vary all the tensors $C$s and $\tilde{C}$s in $W_C^{\rm open}$ and the boundary tensor of $W^{ \rm open }$, $\tilde{A}^1$, $A^2$, $\dots$, $A^b$ and $A^{L-b+1}$, $\cdots$ $A^{L-1}$, $\tilde{A}^{L}$, while fixing the tensors $A^i$ for $b+1\leq i \leq L-b+1$. 

%\commentmb{The last couple of sentences are hard to read. Maybe add some words about what role $b$ plays?}\commentmb{Since the cost function is also $C$, it could get confusing. Can we use a different letter besides $C$?}

Once we find the optimal solution that approximately satisfies Eq. \eqref{loopEqn}, we can cut the closed WLO to create two ends as described in Sec. \ref{sec_Creation} and effectively move the anyon as shown in step IV and V in Fig. \ref{fig_move}.

\subsection{Annihilation of anyon and anti-anyon pairs}

\begin{figure}[b]
    \includegraphics[width=.48\textwidth]{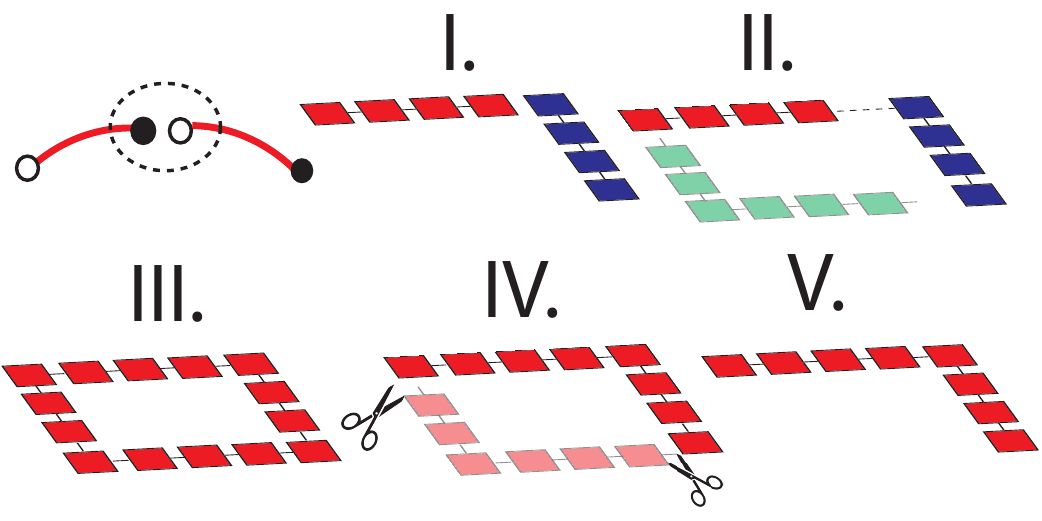}
      \caption{{\bf Annihilation of anyon and anti-anyon pairs} {\bf Step I}: We start with two open WLOs The boundary of the two open WLOs are at nearest neighbor sites. {\bf Step II}: We randomly initialize an open MPO with virtual bonds at the boundary tensors. The two open WLOs and the MPO form a closed MPO. {\bf Step III}: We vary the closed MPO to minimize the cost function defined in Eq. \eqref{cost}. {\bf Step IV and V }: Finally, we cut the closed WLO and keep the tensors on supports of the two open WLOs in Step I.}
      \label{fig_merge}
\end{figure}

In this section, we describe a procedure to fuse an anyon and anti-anyon pair to identity. Given two open WLOs $W_a(\gamma)$ and $W_a(\gamma')$, we show how to join them into a single WLO, as shown in Fig. \ref{fig_merge}, by effectively bringing together two endpoints of $\gamma$ and $\gamma'$ and annihilating the anyons. 

We consider two open WLOs of the form
\begin{eqnarray}
W^{ \rm open }_1 &=& \sum_{\{\alpha\}}A^1_{\alpha_1} A^2_{\alpha_1,\alpha_2} \cdots A^{L_1-1}_{\alpha_{L_1-2},\alpha_{L_1-1}} A^{L_1}_{\alpha_{L_1-1}}, \notag \\
W^{ \rm open }_2 &=& \sum_{\{\beta\}} B^1_{\beta_1} B^2_{\beta_1,\beta_2} \cdots B^{L_2-1}_{\beta_{L_2-2},\beta_{L_2-1}} B^{L_2}_{\beta_{L_2-1}}.
\end{eqnarray}

For simplicity we assume that the tensors $A^{L_1-1}$ and $B^1$ are located at nearest-neighbor sites as shown in step I of Fig. \ref{fig_move}. If this is not the case, we can move the end of $W^{ \rm open }_2$ using the procedure described in Sec. \ref{sec_Moving}.

We then initialize a random MPO with length $L'$ that is a loop complement of $W^{ \rm open }_1$ and $W^{ \rm open }_2$, which takes the form
\begin{eqnarray}
W^{ \rm open }_C = \sum_{\{ \gamma \}} C^1_{\gamma_1} C^2_{\gamma_1,\gamma_2} \cdots C^{L'-1}_{\gamma_{L'-2},\gamma_{L'-1}} C^{L'}_{\gamma_{L'-1}}.
\end{eqnarray}

We can connect the three open WLOs by adding a trivial virtual bond on each boundary tensor as in step III of Fig. \ref{fig_merge}. After adding a virtual bond, the WLOs become
\begin{eqnarray}
\tilde{W}^{ \rm open }_{1, \mu_0, \mu_{L}} &=& \sum_{\{\alpha\}}\tilde{A}^1_{\mu_0, \alpha_1} A^2_{\alpha_1,\alpha_2} \cdots A^{L_1-1}_{\alpha_{L_1-2},\alpha_{L_1-1}} \tilde{A}^{L_1}_{\alpha_{L_1-1}, \mu_{L_1}}, \notag \\
\tilde{W}^{ \rm open }_{2, \nu_0, \nu_{L}} &=& \sum_{\{\beta\}} \tilde{B}^1_{\nu_0,\beta_1} B^2_{\beta_1,\beta_2} \cdots B^{L_2-1}_{\beta_{L_2-2},\beta_{L_2-1}} \tilde{B}^{L_2}_{\beta_{L_2-1}, \nu_{L_2}}, \notag \\
\tilde{W}^{ \rm open }_{C, \sigma_0, \sigma_{L}} &=& \sum_{\{ \gamma \}} \tilde{C}^1_{\sigma_0,\gamma_1} C^2_{\gamma_1,\gamma_2} \cdots C^{L'-1}_{\gamma_{L'-2},\gamma_{L'-1}} \tilde{C}^{L'}_{\gamma_{L'-1}, \sigma_{L'}}.
\end{eqnarray}

We can therefore connect the three WLO and have a closed MPO of the form
\begin{eqnarray}
W = \sum_{\mu_0, \mu_{L_1}, \nu_0, \nu_{L_2}, \sigma_0, \sigma_{L'}} \tilde{W}^{ \rm open }_{1, \mu_0, \mu_{L_1}} 
\tilde{W}^{ \rm open }_{2, \nu_0, \nu_{L_2}}
\tilde{W}^{ \rm open }_{C, \sigma_0, \sigma_{L}}\notag \\
\times\delta_{\mu_{L_1}, \nu_0}
\delta_{\nu_{L_2}, \sigma_0}
\delta_{\sigma_{L'}, \mu_0}.
\end{eqnarray}

To minimize the cost in Eq. \eqref{cost} for the WLO above, we vary the MPO above except for the tensors $A^i_{\alpha_{i-1}, \alpha_i}$, $B^i_{\beta_{i-1}, \beta_i}$, where $b \leq i \leq L_1-b$ . After obtaining a closed WLO as shown in step III of Fig. \ref{fig_merge}, we cut the WLO at the support of $W_C^{ \rm open }$ using the procedure described in Sec. \ref{sec_Creation}. The resulting open WLO is the fusion of $W_1^{ \rm open }$ and $W_2^{ \rm open }$.

\subsection{Topological twist (exchange statistics)}
\label{sec_twist}

\begin{figure*}
    \includegraphics[width=1\textwidth]{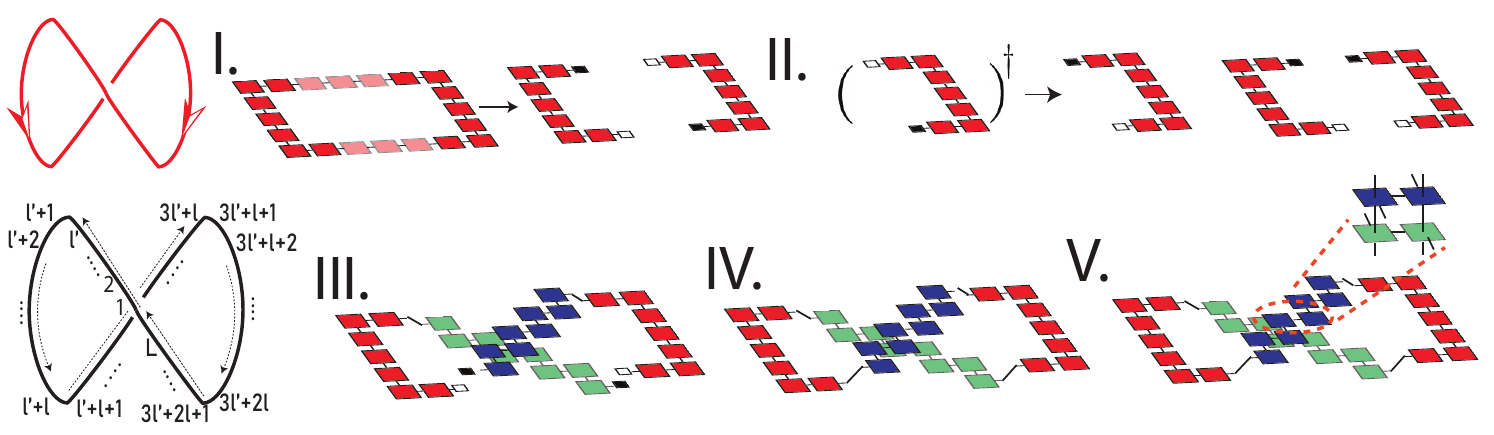}
      \caption{{\bf Topological twist.} {\bf Step I.} We cut a closed WLO and keep two open WLO. {\bf Step II.} We flip the right open WLO to reverse the direction of the anyon and anti-anyon. The boundary sites are denoted as $i$, $j$, $m$ and $n$. {\bf Step III.} We move the anyons at $i$ and $n$ to $m$ and $j$ respectively. {\bf Step IV. } We connect the WLO to form a closed loop. {\bf Step V. } Finally, we contract the physical indices in the self-intersecting region.
      }
      \label{fig_twist}
\end{figure*}

\begin{figure}[h]
    \includegraphics[width=.48\textwidth]{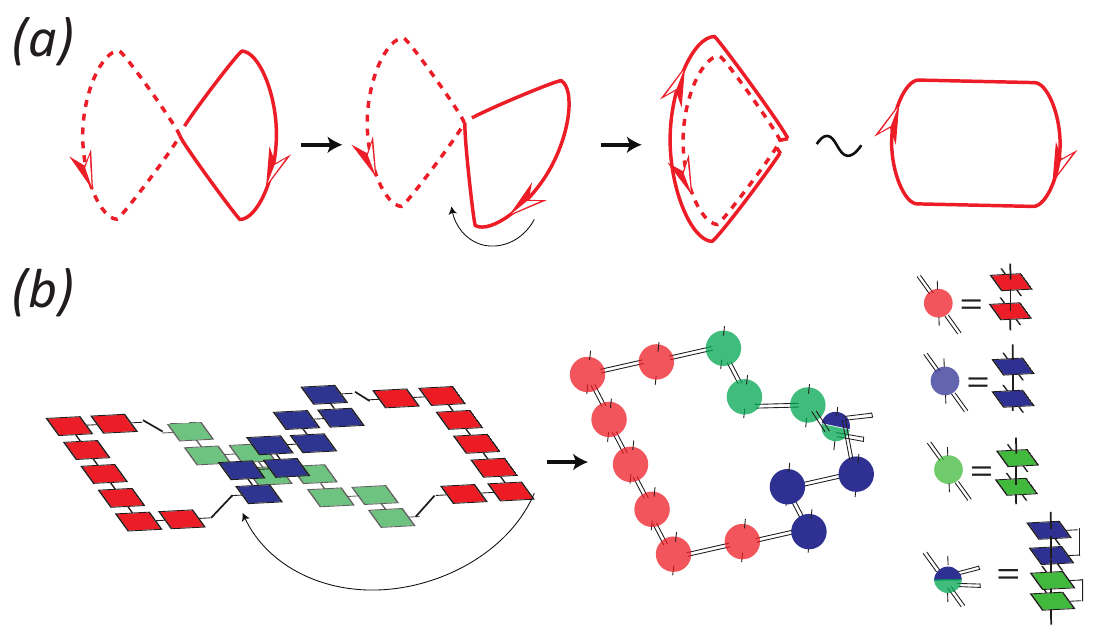}
      \caption{(a) Procedure for creating the support of WLO for process $P_{2}$ defined in Sec. \ref{sec_twist} from process $P_1$. We start with the support of process $P_1$. We rotate the right half of the support and stack it on top of the left half of the support. This WLO is equivalent to a trivial loop. (b) Transformation from the WLO for process $P_1$ to process $P_{2}$, using the procedure described in (a). }
      \label{fig_twist_processII}
\end{figure}

%\commentmb{Does this section assume anything special about the state? Translation invariance, self-dual anyons, reflection symmetry, etc?}

Here we present the scheme that we use to extract the topological twists of the anyons. To calculate the topological twists, we calculate the ratio of the amplitude for the following two processes. In the first process ($P_1$), we create an anyon and anti-anyon pair from the ground state wave function. We then exchange them, and finally, we annihilate the pair of anyon and anti-anyon. In the second process ($P_2$), we create and annihilate the pair directly without any exchange

For the first process, we start with a closed WLO with length $L_o$ of the form 
\begin{eqnarray}
W = \sum_{\{\alpha\}}A^1_{\alpha_0, \alpha_1} A^2_{\alpha_1,\alpha_2} \cdots A^{L_o-1}_{\alpha_{L_o-2},\alpha_{L_o-1}} A^{L_o}_{\alpha_{L_o-1}, \alpha_0}.
\end{eqnarray}

We then cut the closed WLO and keep two segments with length $l$ as shown in step I of Fig. \ref{fig_twist}. These segments represent a creation of an anyon and anti-anyon. The end of the two segments of the WLOs are away from each other by distance $b$ which is an integer larger than $O(\xi)$. 

The WLOs are of the form
\begin{eqnarray}
W^{ \rm open }_1 &=& \sum_{\{\alpha\}}A^1_{\alpha_1} A^2_{\alpha_1,\alpha_2} \cdots A^{l-1}_{\alpha_{l-2},\alpha_{l-1}} A^{l}_{\alpha_{l-1}}, \notag \\
W^{ \rm open }_2 &=& \sum_{\{\alpha\}}A^{b+1}_{\alpha_{b+1}} A^{b+2}_{\alpha_{b+1},\alpha_{b+2}} \cdots A^{b+l-1}_{\alpha_{b+l-2},\alpha_{b+l-1}} A^{b+l}_{\alpha_{b+l-1}}. \notag \\
\label{eq_twist_two_segs}
\end{eqnarray} 

%We then flip the WLO $W^{ \rm open }_2$ to reverse the direction of it as shown in step II of Fig. \ref{fig_twist}.\commentmb{It is not clear what we are assuming about the state that allows us to do this.} Specifically, we swap the tensors in the $W^{ \rm open }_2$ in the following way:
%\begin{eqnarray}
%A^{b+1+i}_{\alpha, \beta} \Longleftrightarrow A^{b+l-i}_{\beta, \alpha},
%\end{eqnarray}
% where the operation $P_{\alpha, \beta} \Longleftrightarrow Q_{\beta, \alpha}$ exchanges the tensor elements $P_{\alpha, \beta}$ and $Q_{\beta, \alpha}$ and  $0 \leq i \leq   \lfloor l/2 \rfloor$. The four sites that support boundary tensors $A^1, A^l, A^{b+1}$, and $A^{b+l}$ are denoted by $i$, $j$, $m$, and $n$ respectively as shown in step II of Fig. \ref{fig_twist}. 

In Step II, we flip the direction of the anyon transport for the WLO $W^{ \rm open }_2$ by applying Hermitian conjugation. 

After cutting and flipping the WLO, we move the anyon located at the open ends $i$ and $n$ to sites $m$ and $j$ respectively, as shown in step III of Fig. \ref{fig_twist}. We then connect the WLOs by annihilating anyon and anti-anyon pairs in step IV of Fig. \ref{fig_twist}. The WLO becomes a self-intersecting closed loop of the form
\begin{eqnarray}
W^{ \rm twist}_{P_1} = \sum_{\{ \beta \}} B^{1}_{\beta_0, \beta_1} B^{2}_{\beta_1, \beta_2} \cdots B^{L-1}_{\beta_{L-2}, \beta_{L-1}} B^{L}_{\beta_{L}, \beta_0},
\label{eq_twisted_W}
\end{eqnarray}
where $L$ is the path length of the closed WLO $W^{ \rm twist}_{P_1}$. The labels for the supports are shown in Fig. \ref{fig_twist}. Finally, we contract the physical indices in the self-intersecting region as shown in step V of Fig. \ref{fig_twist}. 

In step I of process $P_1$, we cut and discard two segments of a closed WLO as described in Eq. \eqref{eq_twist_two_segs}. This step explicitly breaks the gauge symmetry of a matrix product operator, which introduces non-universal complex phases to $W^{\rm open}_1$ and $W^{\rm open}_2$. The non-universal complex phase depends on the details of the implementation. 
In order to cancel the non-universal complex phase, we calculate the amplitude of the second process $P_2$ using the same tensors in Eq. \eqref{eq_twisted_W}, but we contract the tensors without exchanging anyons. We conjecture and numerically verify that the process $P_2$ has the same non-universal complex phase as the process $P_1$. The calculation of process $P_2$ can be achieved by rotating the right half of the support of $W^{ \rm twist}_{P_1}$ around the center of self-intersecting region and stacking it on top of the left half of the support as shown in Fig. \ref{fig_twist_processII}(a). Instead of creating, exchanging and annihilating the anyon and anti-anyon pair, this process creates an anyon and anti-anyon pair and the anyon travels around the left half of the support twice. And subsequently it is annihilated with the anti-anyon. We note that in order to perform the rotation, presumably the ground state wave function must have rotation symmetry around the center of the self-intersecting region and translation symmetry. We have not systematically studied how the above procedure would need to be modified if the translation and rotation symmetries of the system are broken. 

%\commentmb{Why do we have to do this complicated procedure? And are we making any assumptions about the state that allows us to do this?}

The WLO after the rotation is of the following form
\begin{eqnarray}
W^{ \rm twist}_{P2} = \sum_{\{\beta \}}
C_{\beta_1, \beta_{\frac{L}{2}-1}, \beta_{\frac{L}{2}+1}, \beta_{L-1}}
\prod_{i = 2}^{\frac{L}{2}} D^i_{\beta_{i-1},\beta_{L-i+1}, \beta_{i}, \beta_{L-i}}\notag \\
\end{eqnarray}
where 
\begin{eqnarray}
C_{\beta_1, \beta_{\frac{L}{2}-1}, \beta_{\frac{L}{2}+1}, \beta_{L-1}} 
&=& \sum_{\beta_0, \beta_{\frac{L}{2}}} B^{1}_{\beta_0, \beta_1}*
B^{\frac{L}{2}}_{\beta_{\frac{L}{2}-1}, \beta_{\frac{L}{2}}} \notag \\ &*&B^{\frac{L}{2}+1}_{\beta_{\frac{L}{2}}, \beta_{\frac{L}{2}+1}}*
B^{L}_{\beta_{L}, \beta_0},\notag \\
D^i_{\beta_{i-1},\beta_{L-i+1}, \beta_{i}, \beta_{L-i}} &=& B^{i}_{\beta_{i-1}, \beta_i}*B^{L-i+1}_{\beta_{L-i},\beta_{L-i+1}}
\end{eqnarray}
and the tensor multiplication $*$ denotes the contraction of physical indices as shown in Fig. \ref{fig_twist_processII}(b).

Therefore, the topological twist that represents the exchange statistics is calculated from the ratio of the expectation value of $W^{ \rm twist}_{P_1}$ and $W^{ \rm twist}_{P_2}$,
\begin{eqnarray}
\tilde{T}_a = \frac{\langle \psi| W^{ \rm twist}_{P_1;a} | \psi \rangle}{\langle \psi| W^{ \rm twist}_{P_2;a} | \psi \rangle},
\end{eqnarray}
where the label $a$ denotes the anyon type. 

\section{Numerical results}
\label{sec_numeric}
\begin{figure*}[t]
    \includegraphics[width=1\textwidth]{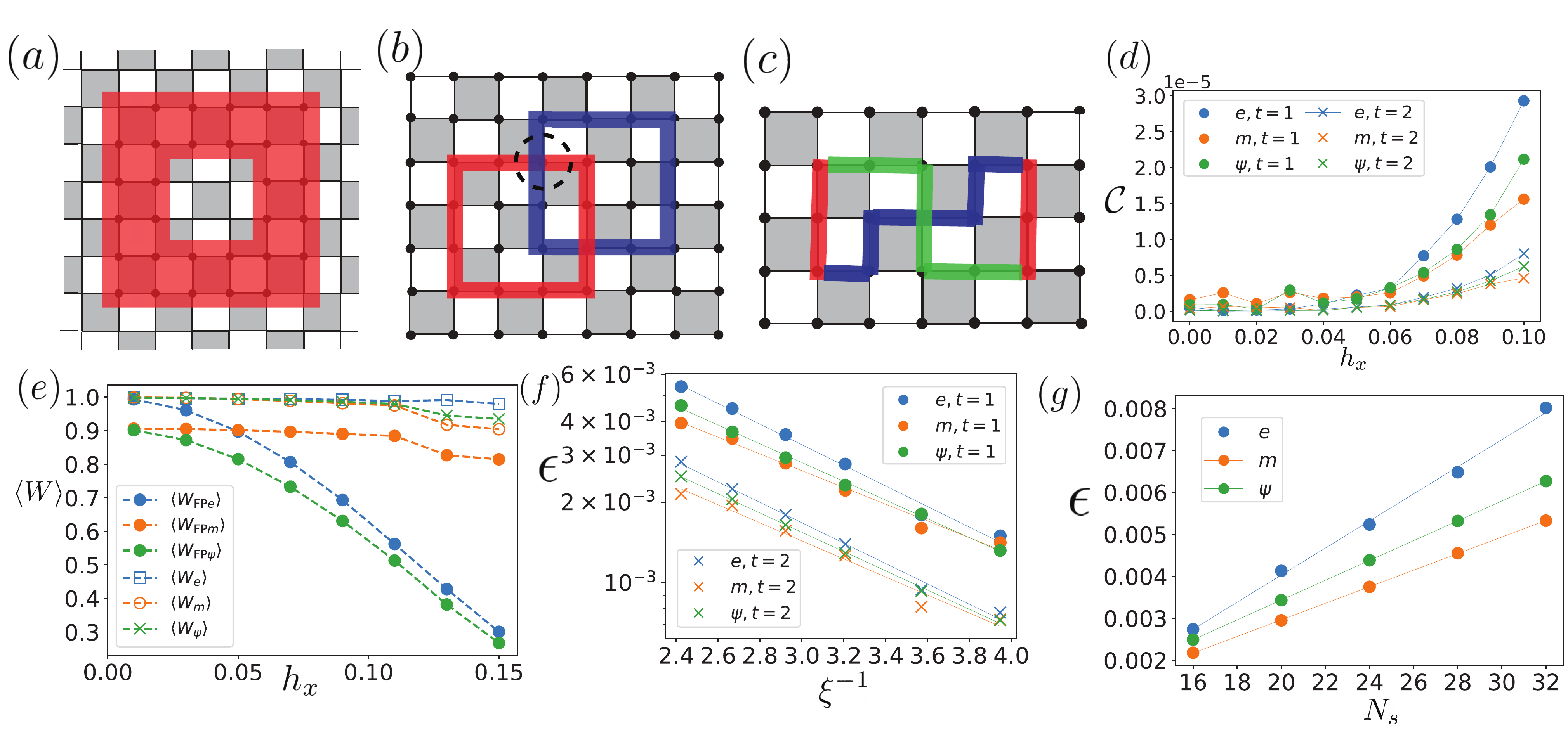}
     \caption{(a) Schematic of toric code model on an infinite plane. The red square is the region supporting the WLO. (b) The red and blue square are regions supporting the WLOs for calculating braiding statistic as described in \ref{sec_Braiding_statistic}. The circled region is region B defined in Eq. \eqref{Twist product}. 
      (c) The curve is the region supporting the WLOs for calculating topological twist as described in \ref{sec_twist}. In the twist region, the blue curve is on top of the green curve. (d) The minimum costs as function of uniform magnetic field $h_x$ for three anyon types in the toric code model with $h_z = 0.05$ and $\chi = 1$. For $t = 1$, the support of the Wilson loops is the perimeter of a   $(L_x,Ly) = (6,4)$ square.  For $t = 2$, the support is depicted in (a). 
      (e) The expectation value of the closed WLOs in renormalization group fixed points ($h_x =h_z = 0$), $W_{\rm FP }$ and the optimized closed WLOs as function of the magnetic field $h_x$. In this figure, $h_z = 0.05$, $L_x = 36$, $L_y = 6$, $\chi = 1$ and $t = 1$.
      (f) The error of the Wilson loop operators $\epsilon = | \langle W - W_{\rm exact} \rangle|$ as function of the inverse of correlation length ($1/\xi$) of the ground state wave function of toric code in uniform magnetic field with $h_z = 0.05$, $0.04 \leq h_x \leq 0.1$, $\chi = 1$.
      (g) The error of the Wilson loop operators $\epsilon$ as function of the number of sites $N_s$ with $h_x = 0.1$, $h_z = 0.05$, $t = 1$, $\chi = 1$, $L_y = 4$. We vary $N_s$ by increasing the side length $L_x$ from $L_x=4$ to $L_x = 8$.}
      \label{fig_tc}
\end{figure*}

In this section, we present the numerical results for extracting WLOs and braiding statistics for various systems.

\subsection{$\mathbb{Z}_2$ toric code model in a magnetic field}

We first consider the $Z_2$ toric code model in a magnetic (Zeeman) field \cite{wu2012phase, vidal2009low, tupitsyn2010topological, halasz2012probing,ritz2021wegner}. The Hamiltonian is of the form
\begin{equation}
    H_{{\rm TC}} = -\sum_{p \in {\rm plaquette}} \hspace{-0.5cm} B_p -\sum_{v \in {\rm vertex}} \hspace{-0.3cm} A_v - \sum_{i} (h_{x}X_i + h_{z}Z_i),
    \label{TC Hamiltonian}
\end{equation}
where the plaquette operators $B_p = \prod_{i \in p} Z_i$, the vertex operators $A_v = \prod_{i \in v} X_i$ and $h_{x}$ and $h_{z}$ are the magnetic fields along the $x$ and $z$ directions respectively.

In the following, we consider a dual lattice, so that the spins are on the lattice sites instead of the bonds. The plaquette operators $A_p$ and the vertice operators $B_v$ of the toric code model are then on alternating plaquettes, as depicted in gray and white respectively in Fig. \ref{fig_tc} (a).

%1. analytical solvable point, correlation length=0, wilson loop has chi=t=1
When the magnetic fields $h_x = h_z = 0$, the ground state of the toric code model can be solved analytically. The ground state has zero correlation length and the Wilson loop operators can be solved exactly (cost $\mathcal{C} = 0$) with bond dimension $\chi = 1$ and thickness $t = 1$ for any size of the closed WLOs $L_x$ and $L_y$. 

There exist four distinct Wilson loop operators for the anyon types : $I$, $e$, $m$ and $\psi$, where $I$ is the identity sector, $e$ and $m$ are bosons with trivial self-braiding phase and a $\pi$ mutual-braiding phase and $\psi = e \times m$ is the fermion. The modular $S$ matrix characterizing the mutual braiding statistics, listing the anyons in the order $I$, $e$, $m$, $\psi$ is of the form,
\begin{eqnarray}
S = \frac{1}{2}
\begin{pmatrix}
1&1&1&1\\
1&1&-1&-1\\
1&-1&1&-1\\
1&-1&-1&1
\end{pmatrix}.
\label{tc_s_matrix_exact}
\end{eqnarray}
The topological twists, which characterize the exchange statistics of the anyons, is given by
\begin{eqnarray}
T = diag(1, 1, 1, -1).
\label{tc_t_matrix_exact}
\end{eqnarray}
Figure \ref{fig_tc} (b) and (c) illustrate the measurement of modular $S$ and $T$ matrices with thickness $t=1$ on 2D square lattice. 

With non-zero magnetic fields, the ground state wave function has a finite correlation length $\xi$. The Wilson loop operators with finite thickness $t$ can no longer represent the exact WLOs. However, the optimization scheme can still find approximate WLOs with error of order $\mathcal{O}(N_s e^{-t/\xi})$, where $N_s$ is the number of sites. 

%\MH{the following several paragraphs can be moved up because they apply to toric and double semion models}
We first obtain the ground state wave function by minimizing the infinite projective entangled pair state (iPEPS) using a recently proposed differential programming approach \cite{liao2019differentiable, crone2020detecting} with corner transfer matrix renormalization group (CTMRG) \cite{orus2009simulation}. 

Next, we follow the protocol described in Sec. \ref{sec_optimization_scheme},  minimizing the cost function defined in Eq. \eqref{cost} starting from random MPO initializations, to find the Wilson loop operators. Figure \ref{fig_tc}(d) shows the minimum cost as a function of $h_x$ with a fixed $h_z = 0.05$ for thickness $t = 1$ and $t = 2$. The cost decreases when the thickess $t$ increases, demonstrating that a larger thickness gives a better approximation to the true Wilson loop operator. 

In Fig. \ref{fig_tc}(e), we compare the optimized closed WLOs to the WLOs known for the fixed point toric code Hamiltonian ($h_x = h_z = 0$). The expectation values of the optimized closed WLOs stay close to $1$ as the magnetic field is increased. However, the expectation values of the fixed point WLOs decrease with increasing $h_x$. We note that this is remarkable given that we do not use prior knowldge of the WLOs of the fixed point Hamiltonian; our scheme is completely unbiased and uses no prior information aside from the ground state of the perturbed Hamiltonian whose WLOs we are trying to find.  

While we do not know the form of the exact WLOs in the presence of non-zero $h_x$, $h_z$, we do know that their expectation value in the ground state is 1. Therefore, we can compute how close our optimized WLOs are to the exact WLOs. We define the error
\begin{align}
\epsilon = |\langle \psi | W - W_{\rm exact}|\psi\rangle |,
\end{align}
where we use the fact that the exact WLO, $W_{\rm exact}$, satisfies $\langle W_{\rm exact} \rangle = 1$. Here $W$ represents a WLO found through our optimization protocol. 

%\commentmb{How do you numerically extract $\xi$?} 
We present the error $\epsilon$ of the WLOs as a function of the inverse of the correlation length $1/\xi$ and number of sites $N_s$ in Fig. \ref{fig_tc}(f) and (g), respectively The correlation length is computed by through the exponential decay of the correlation funcion, $\langle X_{x_0, y_0} X_{x_0+d, y_0}\rangle - \langle X_{x_0, y_0}\rangle \langle X_{x_0+d, y_0}\rangle = Ae^{-d/\xi} $. The largest correlation length we reach in our simulation is $\xi = 1.21$.

%\commentmb{Doesn't it matter how many optimization steps were taken? And is this an average over many runs or what? How many samples are chosen for the average? } \ZP{do we need to list the number of optimization steps for all the optimized WLOs? I mentioned in Sec. III that we need a few hundred or a few thousand iterations to reach minimum. Also the number of sample is fixed throughout the manuscript as defined in Sec. III. } 
In Fig. \ref{fig_tc}(f), we vary the correlation length by varying the magnetic field $h_x$. We show that the error drops exponentially as function of $1/\xi$. Fig. \ref{fig_tc}
(g) shows how the error $\epsilon$ scales with the number of sites $N_s$ in the support of the WLO with $t = 1$, indicating that the error scales up linearly with $N_s$. Therefore the error scaling is consistent with the error bound $\mathcal{O}(N_s e^{-t/\xi})$ \cite{hastings2005quasiadiabatic}. %\commentmb{This is actually confusing because I thought that increasing the size of the WLO should actually improve its behavior.} \ZP{it improves since error goes down exponentially with thickness. But there is also a prefactor $N_s$ in the error bound.}

Finally, we numerically evaluate the twist product matrix $\tilde{S}_{i,j}$ and the topological twist $\tilde{T}_i$. The twist product matrix is consistent with Eq. \eqref{tc_s_matrix_exact} up to $10^{-5}$ error and the topological twist has error up to $10^{-2}$. For example, the twist product matrix for $h_x = 0.1$, $h_z = 0.05$, $t = 2$, $L_x = 6$, $L_y=4$, $\chi = 1$ is 
\begin{eqnarray}
\tilde{S} = 
\begin{pmatrix}
1.0 & 1.0 & 1.0 & 1.0 \\
1.0 & 1.0 & -1.00001 & -0.99996 \\
1.0 & -0.99999 & 1.0 & -0.99977 \\
1.0 & -1.00004 & -1.00024 & 1.0 \\
\end{pmatrix},
\end{eqnarray}
and
\begin{eqnarray}
\tilde{T} = diag(1.0, 0.963, 0.912, -0.974).
\end{eqnarray}

\subsection{Double semion model}
\begin{figure}[h]
    \includegraphics[width=.48\textwidth]{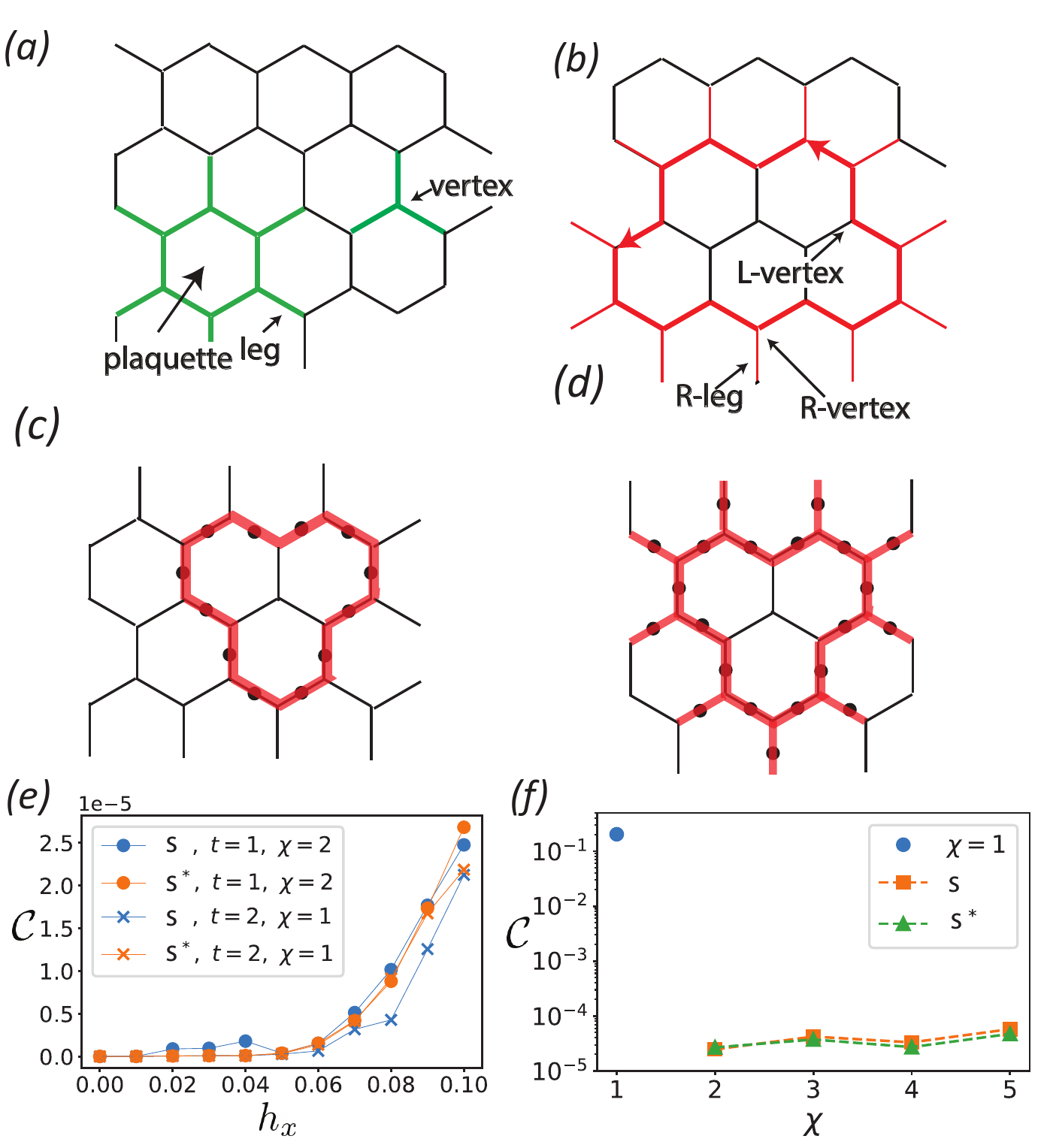}
      \caption{(a) Schematic of the plaquette and vertex for double semion model. The ``legs" of a plaquette are the edges that stick out of the plaquette. (b) Schematic of the L-vertex, R-vertex and R-leg for double semion model. (c) and (d) The support of the Wilson loop operators with thickness $t = 1$ and $t = 2$ respectively.  
      (e) The minimum cost as a function of uniform magnetic field $h_x$ for two semions in the double semion model. The support of the Wilson loops are depicted in (c) and (d). 
      (f) The minimum cost as a function of the bond dimension $\chi$ for the two semions for $h_x = 0.1$ and $t = 1$}
      \label{fig_ds}
\end{figure}
In this section, we present our numerical results for the double semion model. 

The Hamiltonian of the double semion model \cite{levin2005string}, which we take to be on the honeycomb lattice, is of the form
\begin{eqnarray}
H_{DS} &=& -\sum_{v \in vert.} A_v - \sum_{i} h_{x}X_i, \notag \\
 &+& (\sum_{p \in plaq. }B_p \prod_{j \in \rm{legs}(p)}i^{\frac{1-X_j}{2}} + h.c.)
\label{DS_hamiltonain}
\end{eqnarray}
where the plaquette operators $B_p = \prod_{i \in p} Z_i$, the vertex operators $A_v = \prod_{i \in v} X_i$ and legs$(p)$ is the legs of plaquette $p$ as shown in Fig. \ref{fig_ds}(a). 
%the Hilbert space spanned by the closed string configurations, which satisfy $B_v|$close string$\rangle = |$close string$\rangle$, $\forall v$ \commentmb{This could be explained better. In what basis are the strings defined. Are the states on the lhs and rhs different or same states?}. To preserve the closed string configuration in the low-energy subspace, we only apply a magnetic field along the $x$ direction. \commentmb{I don't understand this. Isn't our method supposed to work in general?}

For $h_x = 0$, the ground state and the Wilson loop operators can be obtained analytically \cite{levin2005string}. There exists four distinct WLOs denoted by $I$, $s$,  $s'$, $b = s \times s'$, which represent identity, right and left-handed semions and a boson. %\commentmb{semions are usually labeled $s$, $s'$ and the boson is $b = s \times s'$.} 
The two semions have $\pi$ self-braiding statistics and trivial mutual braiding statistics. The modular $S$ matrix, listing the anyons in the order $I$, $s$,  $s'$, $b$, is of the form
\begin{eqnarray}
S = \frac{1}{2}
\begin{pmatrix}
1&1&1&1\\
1&-1&1&-1\\
1&1&-1&-1\\
1&-1&-1&1
\end{pmatrix}.
\label{ds_s_matrix_exact}
\end{eqnarray}
The topological twists that characterize the exchnage statistics of the double semion model is given by 
\begin{eqnarray}
T = diag(1, i, -i, 1).
\end{eqnarray}

The exact closed WLOs for $h_x = 0$ are of the form
\begin{eqnarray}
W_I &=& I, \notag \\
W_{s } &=& \prod_{l \in \gamma} Z_l \prod_{k \in \rm{L-vertex}} \hspace{-0.3cm} (-1)^{\frac{1}{4}(1-X_i)(1+X_j)} \prod_{l \in \rm{R-leg}} \hspace{-0.25cm} i ^{\frac{1}{2}(1-X_l)} \notag \\
W_{s' } &=& \prod_{l \in \gamma} Z_l \prod_{k \in \rm{L-vertex}} \hspace{-0.3cm} (-1)^{\frac{1}{4}(1-X_i)(1+X_j)} \prod_{l \in \rm{R-leg}} \hspace{-0.25cm} (-i) ^{\frac{1}{2}(1-X_l)} \notag \\
W_{b} &=& \prod_{l \in \mathcal{\gamma}'}X_l ,
\label{Wilson_exact_DS}
\end{eqnarray}
where $\mathcal{\gamma}$ is a path of a closed WLO, $\gamma'$ is a path of a closed WLO on the duel lattice, $i$ and $j$ are two legs attached to L-vertex $k$ as shown in Fig. \ref{fig_ds}(b). 

Note that the $W_s$ and $W_{s'}$ operators above have thickness $t = 2$ and bond dimension $\chi = 1$. Interestingly, using our unbiased numerical optimization, we found an equivalent way to represent the exact WLOs at the fixed point with $t=1$ and $\chi = 2$. We present its analytical form in Appendix \ref{sec_alternative_ds_wlo}.%\ZP{mention appendix A}

In order to numerically optimize the WLOs, we numerically optimize the iPEPS ground state wave function \cite{corboz2012simplex} and consider WLOs with thickness $t = 1$ and $t = 2$ as shown in Fig. \ref{fig_ds}(c) and \ref{fig_ds}(d) respectively. Fig. \ref{fig_ds}(e) shows the minimum cost achieved as a function of the magnetic field $h_x$. The optimizer converges to these minimum costs after roughly 1000 iterations. Fig. \ref{fig_ds}(f) shows the minimum cost as a function of the bond dimension $\chi$ for the two semions for $h_x = 0.1$ and $t = 1$. 

%In addition to the solution with $t = 2$ and $\chi = 1$, we show that there exists another set of solutions with thickness $t = 1$ and bond dimension $\chi = 2$, and the minimum cost of the two sets of solutions are of similar order. We further increase the bond dimension $\chi$. However, the cost does not change significantly for $\chi \geq 2$ as shown in Fig. \ref{fig_ds}(f). To further understand the results with $t=1$ and $\chi \geq 2$, in the Appendix \ref{sec_alternative_ds_wlo}, we analytically show that when $h_x = 0$, one can have equivalent WLOs to the Eq. \eqref{Wilson_exact_DS} with $t=1$ and $\chi = 2$.
%\commentmb{I find the preceding paragraph hard to understand. }

\begin{figure}[t]
    \includegraphics[width=.48\textwidth]{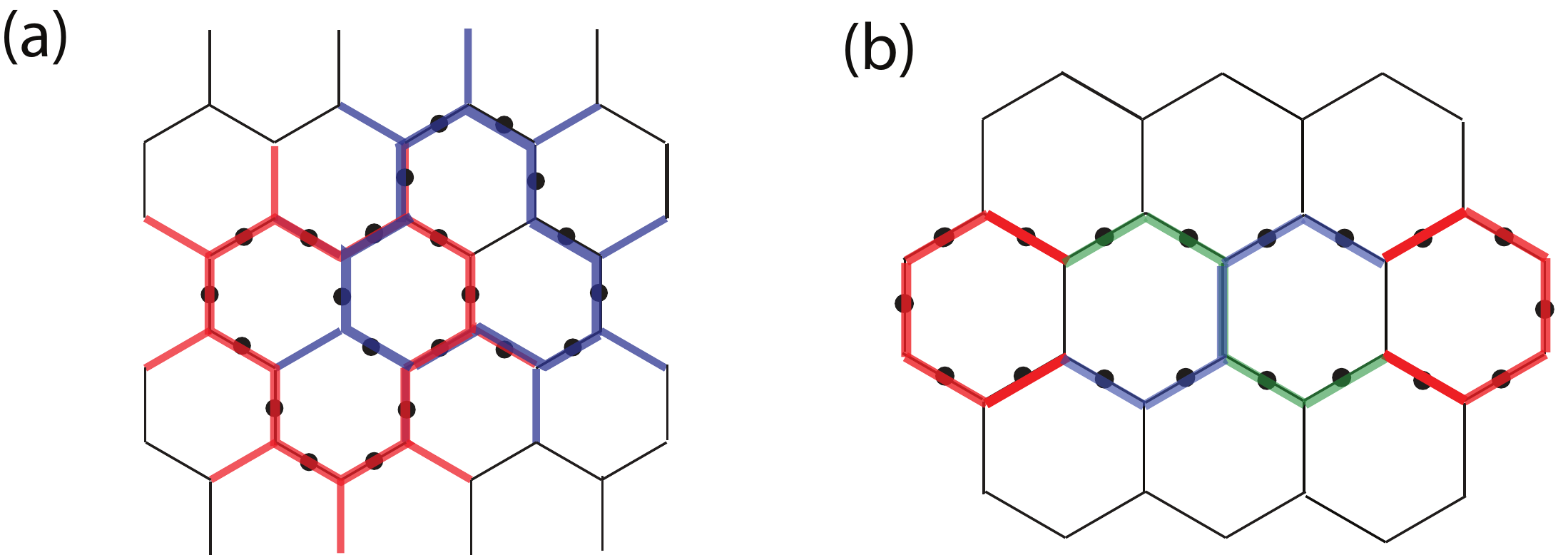}
      \caption{(a) The red and blue curves are regions supporting the WLOs for calculating braiding statistic for double semion model. (b) The twisted curve is the regions supporting the WLOs for calculating topological twist as described in \ref{sec_twist}. In the overlaping region, the blue curve is on top of the green curve.  }
      \label{fig_ds_s_t_anyon}
\end{figure}

%Finally, we present the numerical results for manipulation of anyon, mutual- and self- braiding statistic for double semion model. 

We then numerically evaluate the twist product matrix $\tilde{S}_{i,j}$ and the topological twist $\tilde{T}_i$. The support of $\tilde{S}$ and $\tilde{T}$ are shown in Fig. \ref{fig_ds_s_t_anyon}(a) and (b) respectively. The twist product matrix is consistent with Eq. \eqref{ds_s_matrix_exact} up to $10^{-4}$ error and the topological twist has error up to $10^{-2}$. For example, the twist product matrix for $h_x = 0.1$, $t = 1$, $\chi = 2$ is
\begin{eqnarray}
\tilde{S} = 
\begin{pmatrix}
1.0&1.0&1.0&1.0\\
1.0&-1.0&1.001&-0.994\\
1.0&0.999&-1.0&-1.002\\
1.0&-1.006&-0.998&1.0
\end{pmatrix},
\end{eqnarray}
and 
\begin{eqnarray}
\tilde{T} = {\rm diag}(1.0, 0.9351e^{i \pi 0.502}, 0.9412e^{-i \pi 0.489}, 0.993).
\end{eqnarray}
The correlation length of the ground state, calculated through the exponential decay of the correlation funcion, $\langle X_{x_0, y_0} X_{x_0+d, y_0}\rangle - \langle X_{x_0, y_0}\rangle \langle X_{x_0+d, y_0}\rangle$, with $h_x = 0.1$ is $\xi = 0.87$.

%Complexity : 
%\begin{itemize}
%    \item computational cost goes up exponentially with thickness, which is proportional to correlation length
%    \item bond dimension : the alternative optimization favors WLO operator with small bond dimension. It makes the unbias search for WLO with large bond dimension difficult.
%\end{itemize}

\section{Summary and outlook}
\label{sec_summary_and_outlook}

In conclusion, we propose a numerical optimization-based scheme to extract Wilson loop operators of anyons from a single ground state wave function defined on a simply connected region of space. We show how, after extracting closed loop operators, one can then modify them to obtain Wilson line operators that create, move, and annihilate anyons. This allows us to ultimately extract the braiding statistics and topological twists of the anyons from a single bulk ground state wave function. While our protocol for extracting the modular $S$ matrix is expected to be general, our protocol for extracting the topological twists  may be benefiting from the lattice symmetries of our models; we leave a systematic investigation of this for future work. 

Our algorithm fully succeeds only when all distinct equivalence classes of Wilson loop operators have been found. We expect that in general, for a large enough bond dimension, all Wilson loop operators can be captured by our matrix product operator ansatz. In practice it may be the case that some Wilson loop operators might be more complicated than others, in the sense of having higher operator entanglement or requiring larger bond dimension. In this case, if there is an implicit bias of the optimization procedure towards the simpler loop operators, then the algorithm may never succeed in discovering a complete set of loop operators starting from random initialization. In this case, further work needs to be done on either finding improved initializations or  modifying the optimization procedure to remove the implicit biases. 

Our work also raises an interesting question of whether all Wilson loop operators for anyons can always be described by an MPO with finite bond dimension. This is particularly intriguing to study for chiral topological orders, such as fractional quantum Hall states, where loop operators have never been explicitly computed in MPO form and the ground state wave functions cannot be described by a PEPS with finite bond dimension. 

%However, we note that the algorithm succeeds only when all of the distinct loop operators for a ground state wave function can be approximated by MPOs with finite and small bond dimension. It leads to an open question how large a bond dimension we need in order to approximate any possible WLO to certain accuracy. It is particularly intriguing to study the WLO for the chiral topological order such as the fractional Chern insulator.

%
%\ZP{It is widely believed that the matrix product operators can represent any arbitrary operator with a sufficiently large bond dimension. However, even if we use a sufficiently large bond dimension in the WLO ansatz, the numerical optimization procedure does not guarantee to reach the full set of WLOs since the numerical optimization procedure implemented in this work favors the optimal solution with a low bond dimension, i.e., identity operator. This bias could prohibit us from finding the full set of WLO if some WLOs have a significantly higher bond dimension than others, such as the double Fibonacci string net model \cite{bultinck2017anyons}. Therefore, it is important to explore other optimization schemes and different WLO ansatz in order to reduce the bias in the optimization scheme.}

So far, our optimization scheme is tailored to Abelian topological orders. It would be interesting to generalize it to the case of non-abelian topologically ordered phases, which can have anyons with quantum dimension greater than one. One possible direction is to design an optimization scheme to extract both the WLOs and the quantum dimension of anyons simultaneously.

As we discussed, the Wilson line operators can be generalized to more generic movement and splitting operators. If we discover such generic movement and splitting operators through a similar optimization approach to what we have described, it may also be possible to extract the $F$ and $R$ symbols of the underlying modular tensor category using the results of Ref. \cite{kawagoe2020microscopic}. 

Looking further, similar optimization procedures applied to symmetry defect creation and movement operators may eventually allow us to extract the full $G$-crossed braided tensor category \cite{barkeshli2019} that describes a given symmetry-enriched topological ground state. This would allow the extraction of all possible topological invariants from a single bulk ground state wave function. 

%From a fundamental perspective, it is intriguing to ask if, given any wave function with short-range correlations and area law entanglement, a complete set of data for topological order can be obtained. In particular, this implies removing the guarantee that the wave function be a ground state of a local, gapped Hamiltonian. 

%However, there exist models that are believed to be gapless, and the ground state wave function fails to exhibit several features of topological order. Yet, they have loop operators satisfying Eq. \eqref{loopEqn} and the twist products of loop operators are not equal to one. Examples include the two-dimensional quantum compass model (Bacon-Shor's code) \cite{dorier2005quantum, bacon2006operator} and the deformed toric code model \cite{castelnovo2008quantum}. It is therefore interesting to further investigate the locally invisible operator \cite{haah2016invariant} in the gapless system.

More broadly, recent developments in quantum simulators allow the realization of topologically ordered states that might not occur in conventional electronic matter \cite{Googletoric2021,HarvardQSL1}. Given this opportunity, it is intriguing to develop measurement protocols for probing topological properties of a ground state wave function associated with a prior unknown gapped Hamiltonian. Our optimization protocol may be particularly relevant in this context. Since our scheme requires 
measuring several observables for a given wave function, it may be useful when combined with shadow tomography. \cite{huang2020predicting}.

\section{Acknowledgements}

This work is supported by NSF CAREER DMR- 1753240 (MB), AFOSR-MURI FA9550- 19-1-0399, ARO W911NF2010232, ONR N00014-20-1-2325, National Science Foundation QLCI grant OMA-2120757, U.S. Department of Energy, Quantum Systems Accelerator program and the Simons and Minta Martin Foundations (MH,ZC).

\appendix
\section{Alternative definition of Wilson loop operator for double semion model}
\label{sec_alternative_ds_wlo}
\begin{figure}[h]
    \includegraphics[width=.48\textwidth]{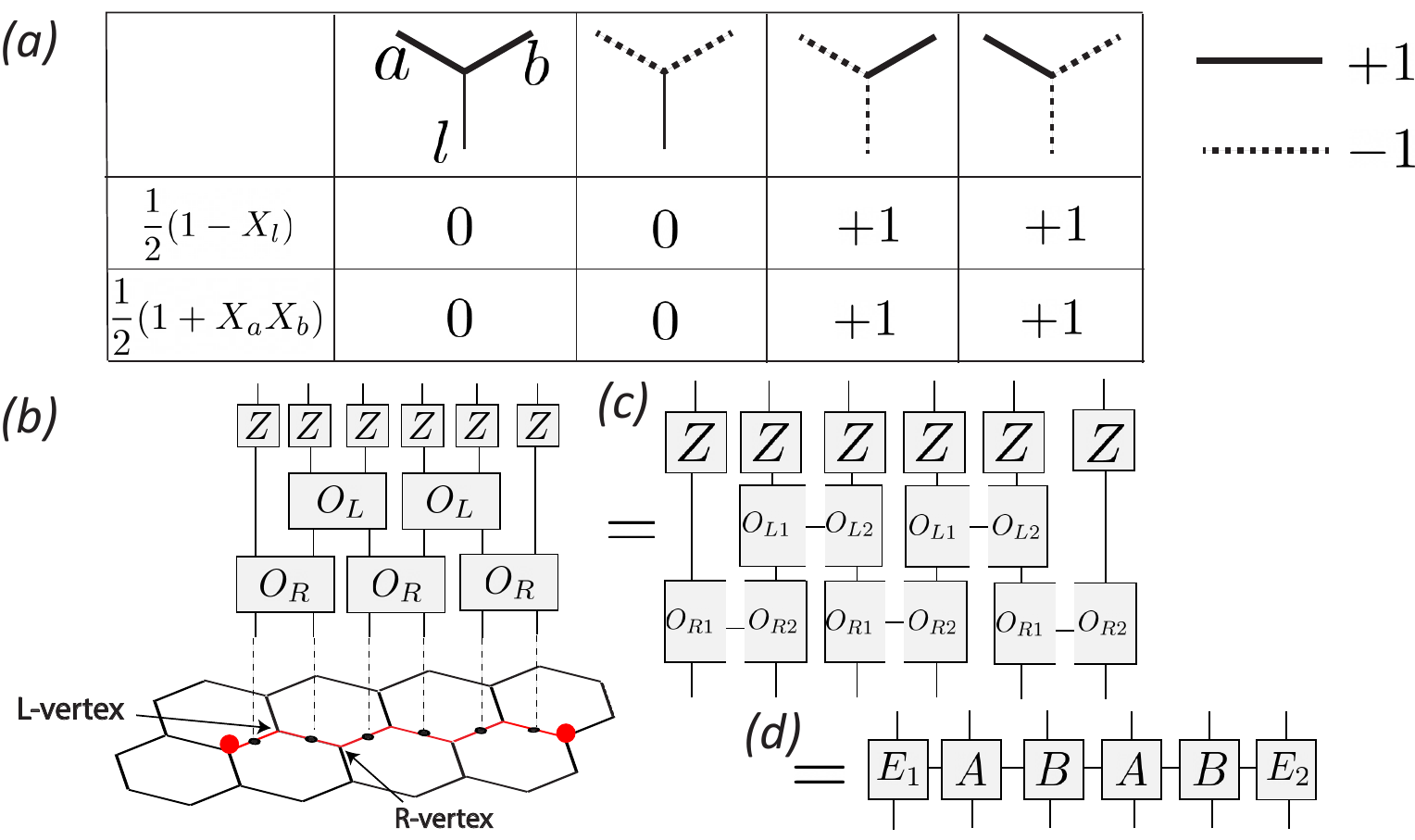}
      \caption{Alternative representation of $W_s$ and $W_{s'}$.}
      \label{fig_ds_MPO}
\end{figure}

In Eq. \eqref{Wilson_exact_DS}, the operators applied on the R-leg $l$ are $(\pm i)^{\frac{1}{2}(1-X_l)}$. The operator in the exponent $\frac{1}{2}(1-X_l)$ is equivalent to $\frac{1}{2}(1+X_aX_b)$ in the space spanned by close string configurations as shown in fig. \ref{fig_ds_MPO}(a), where $a$ and $b$ are the two sites attached to the same vertex of the R-leg $l$. Therefore, the Wilson loop operators can be rewritten as 

\begin{eqnarray}
W_{s } &=& \prod_{i \in \gamma} Z_i \prod_{k \in L} (-1)^{\frac{1}{4}(1-X_i)(1+X_j)} \prod_{l \in R} i ^{\frac{1}{2}(1-X_aX_b)} \notag \\
W_{s' } &=& \prod_{i \in \gamma} Z_i \prod_{k \in L} (-1)^{\frac{1}{4}(1-X_i)(1+X_j)} \prod_{l \in R}  (-i) ^{\frac{1}{2}(1-X_aX_b)}, \notag \\
\label{Wilson_DS_alternative}
\end{eqnarray}
where $\mathcal{\gamma}$ is a path of a closed WLO, $R$ denotes R-vertex and $a$ and $b$ are two legs attached to R-vertex $l$ as shown in fig. \ref{fig_ds}(b).

To show that the WLOs in Eq. \eqref{Wilson_DS_alternative} can be represented by MPOs with $t = 1$, $\chi = 2$, we consider an open Wilson operator shown in fig. \ref{fig_ds_MPO}(b) which creates semions on the both ends of the Wilson operator for illustration. A closed WLO can be defined in a similar fashion. To simplify the notation, we define $O_R = (\pm i) ^{\frac{1}{2}(1-X_aX_b})$ for right and left-handed semion respectively and $O_L = (-1)^{\frac{1}{4}(1-X_i)(1+X_j)}$. The Wilson loop operators can be represented in the tensor network notation as shown in \ref{fig_ds_MPO}(b). The operators $O_R$ and $O_L$ can be decomposed through singular value decomposition (SVD) as 
\begin{eqnarray}
O_R = \sum_{\alpha=1}^2 O_{R1, \alpha} O_{R_2, \alpha } \notag \\
O_L = \sum_{\alpha=1}^2 O_{L1, \alpha} O_{L_2, \alpha },
\end{eqnarray}
where $\alpha$ is the index auxiliary bond as shown in fig. \ref{fig_ds_MPO}(c) and 
\begin{eqnarray}
O_{R_1} &=& 
\sqrt{\frac{i}{2}} 
\begin{pmatrix}
I &
-iX
\end{pmatrix}, ~
O_{R_2} = 
\begin{pmatrix}
I \\
X
\end{pmatrix} \notag \\
O_{L_1} &=& 
\begin{pmatrix}
I &
-2P_{-}
\end{pmatrix}, ~
O_{L_2} =
\begin{pmatrix}
I \\
P_{+}
\end{pmatrix},
\end{eqnarray}
where $P_{\pm} = \frac{1}{2}(1\pm X)$. After SVD, the Wilson operators can be expressed by a two-site periodic MPO terminated by edge tensors :
\begin{eqnarray}
W_{s} = \sum_{\{\alpha\}} E_{1 \alpha_{1}} A_{\alpha_1 \alpha_2} B_{\alpha_2 \alpha_3} \dots A_{\alpha_{L-2} \alpha_{L-1}} B_{\alpha_{L-1} \alpha_L} E_{2 \alpha_L},\notag \\ 
\end{eqnarray}
where $A_{\alpha, \beta} = Z O_{L_1, \beta} O_{R_2, \alpha}$, 
$B_{\alpha, \beta} = Z O_{L_2, \alpha} O_{R_1, \beta}$, 
$E_{1(2)} = ZO_{R_{1(2)}}$ and $\alpha, \beta = 1,2$. $W_{s'}$ can be defined similarly. Therefore, the WLO for creating semions can be expressed by MPO with $t = 1$ and $\chi = 2$.

\bibliographystyle{unsrt}

\clearpage

\end{document}